\documentclass[aps,prb,superscriptaddress,longbibliography,twocolumn]{revtex4-2}

\usepackage[en-US,useregional,showdow]{datetime2}
\DTMlangsetup[en-US]{ord=raise}

\usepackage{amsthm}
\theoremstyle{definition}

\theoremstyle{plain}

\usepackage{graphicx}
\usepackage{longtable}
\usepackage{wrapfig}
\usepackage{rotating}
\usepackage[normalem]{ulem}
\usepackage{amsmath}
\usepackage{amssymb}
\usepackage{capt-of}
\usepackage{hyperref}
\usepackage{amsthm}
\usepackage{xcolor}
\usepackage{physics}
\usepackage{slashed}
\usepackage{siunitx}
\usepackage{mathtools}
\usepackage{bm}
\usepackage{bbm}
\usepackage[capitalize]{cleveref}
\Crefname{equation}{Eq.}{Eqs.}

\usepackage[inline]{enumitem}
\newlist{captionlist}{enumerate*}{2}
\setlist[captionlist,1]{label=\textbf{(\alph*)}}
\setlist[captionlist,2]{label=\textbf{(\alph{captionlisti}.\roman*)}}

\let\oldsout\sout
\renewcommand{\sout}[1]{{\color{red}{\oldsout{#1}}}}

\newcounter{para}
\newcommand{\para}{\par\refstepcounter{para}\textbf{{\color{cyan}[\thepara]}}\space}
\let\para\relax

\newcommand{\xCornell}{Department of Physics, Cornell University, Ithaca, NY, USA}
\newcommand{\xEwha}{Department of Physics, Ewha Womans University, Seoul, South Korea}
\newcommand{\xHarvard}{Department of Physics, Harvard University, Cambridge, MA, USA}
\newcommand{\xRadcliffe}{Radcliffe Institute for Advanced Studies, Cambridge, MA, USA}

\begin{document}
\author{Haining Pan}
\affiliation{\xCornell}
\author{Eun-Ah Kim}
\affiliation{\xCornell}
\affiliation{\xRadcliffe}
\affiliation{\xHarvard}
\affiliation{\xEwha}
\author{Chao-Ming Jian}
\affiliation{\xCornell}

\date{\today}
\title{Realizing a tunable honeycomb lattice in ABBA-stacked twisted double bilayer WSe$_2$}
\begin{abstract}

The ideal honeycomb lattice, featuring sublattice and SU(2) spin rotation symmetries, is a fundamental model for investigating quantum matters with topology and correlations.
With the rise of the moir\'e-based design of model systems, realizing a tunable and symmetric honeycomb lattice system with a narrow bandwidth can open access to new phases and insights. 
We propose the ABBA-stacked twisted double bilayer WSe$_2$ as a realistic and tunable platform for reaching this goal.
Adjusting the twist angle allows the bandwidth and the ratio between hopping parameters of different ranges to be tuned. Moreover, the system's small bandwidth and spin rotation symmetry enable effective control of the electronic structure through an in-plane magnetic field. We construct an extended Hubbard model for the system to demonstrate this tunability and explore possible ordered phases using the Hartree-Fock approximation.
We find that at a hole filling of $\nu = 2$ (two holes per moiré unit cell), an in-plane magnetic field of a few Tesla can ``dope" the system from a semimetal to a metal. Interactions then drive an instability towards a canted antiferromagnetic insulator ground state. Additionally, we observe a competing insulating phase with sublattice charge polarization. Finally, we discuss the experimental signatures of these novel insulating phases.
\end{abstract}

\maketitle

\section{Introduction}

\para 
The ideal honeycomb lattice with both sublattice and SU(2) spin rotation symmetries is an important platform for studying many-body physics with topology and electronic correlations.  Graphene has been widely studied as a material realization of this lattice. Yet, the system's wide bandwidth limits the exploration of the full phase space of honeycomb lattice systems. Realizing a symmetric honeycomb lattice with a narrow bandwidth and enhanced tunability would significantly expand the accessible phase space, enabling the exploration of higher doping levels and stronger correlations. Among other possibilities, this advancement would facilitate experimental investigations of various intriguing correlated phases proposed earlier, such as charge and spin density waves~\cite{honerkamp2008density,aleiner2007spontaneous,bercx2009magnetic}, chiral superconductivity~\cite{nandkishore2012chiral,nandkishore2014superconductivity}, and topological Mott insulators~\cite{raghu2008topological}. Over the past few years, moir\'e superlattices in transition metal dichalcogenides (TMD) multilayer structures have emerged as a highly-tunable platform for quantum simulation of two-dimensional lattice models with narrow bandwidth~\cite{regan2020mott,tang2020simulation,xu2020correlated,wang2020correlated,ghiotto2021quantum,zhou2021bilayer,li2021imaging,li2021quantum,gu2021dipolar,zhao2022realization,xie2022valleypolarized,xie2022topological,tao2022valleycoherent,liu2021excitonic,foutty2023mapping}. Triangular superlattices~\cite{wang2020correlated,regan2020mott,tang2020simulation,xu2020correlated,zhou2021bilayer,li2021imaging,li2021quantum,ghiotto2021quantum} and honeycomb superlattices with spin-orbit coupling (SOC) or asymmetric sublattices~\cite{li2021quantum,tao2022valleycoherent,foutty2023mapping,zhang2021spintextured,wu2019topological} have been realized. In this study, we aim to find a twisted TMD multilayer structure that realizes an ideal honeycomb lattice with both sublattice and SU(2) spin rotation symmetries.

\para 
Theoretically, a strategy to create an ideal emergent honeycomb lattice with sublattice and SU(2) spin symmetries in TMD multilayer structures follows three criteria as first pointed out in Refs.~\cite{angeli2021valley,zhang2021electronic,xian2021realization}: 
(1) The $\Gamma$-valley of each TMD layer, where SOC is negligible, appears at the valence band edge (rather than the $K$/$K'$-valleys), forming an emergent lattice for the doped holes~\cite{angeli2021valley,zhang2021electronic,xian2021realization,movva2018tunable}. 
(2) A symmetry relates the independent high-symmetry-stacking MX and XM sites [see Fig.~\ref{fig:schematic}(b)], which form the two sublattices of the emergent honeycomb lattice; 
(3) The energy at the MX and XM sites is higher than the MM site (another high-symmetry stacking location)~\cite{pan2020band}. However, previous papers only proposed these strategies in TMD bilayers. The correct energetics (in criteria (1) and (3)) in these proposals are yet to be confirmed and realized by experiments. For example, Refs.~\onlinecite{gatti2022observationa,pei2022observation} have experimentally identified the $\Gamma$-valley physics in twisted bilayer WSe$_2$ but found it to appear in energies much below the valence band edge, which is located at the $K$/$K'$-valleys. 

Here, we pursue realizing an ideal emergent honeycomb lattice in ABBA-stacked twisted double bilayer WSe$_2$ (tdbWSe$_2$), i.e., two AB- and BA-stacked WSe$_2$ bilayers with a small relative twist angle [see Fig.~\ref{fig:schematic}(a)]. As experimentally shown in a closely related system, the ABAB-stacked tdbWSe$_2$, the increase in the number of layers enhances the interlayer hybridization driving the $\Gamma$-valley to the valence band edge \cite{movva2018tunable, foutty2022tunable}. However, the ABAB-stacked tdbWSe$_2$ does not have the symmetry needed for criterion (2) and only realizes a $\Gamma$-valley-based emergent triangular lattice. In this paper, we show that the ABBA-stacked tdbWSe$_2$ can pass the three criteria and realize a tunable ideal honeycomb lattice with both sublattice and SU(2) spin rotation symmetries.

\para 
The remainder of this paper is organized as follows. In Sec.~\ref{sec:theory}, we provide the microscopic analysis that the ABBA-stacked tdbWSe$_2$ can pass the three criteria above and, hence, leads to an emergent honeycomb lattice with both sublattice and SU(2) spin rotation symmetries. We construct a continuum model for this tdbWSe$_2$ system, enabling estimations of the hopping parameters of the emergent honeycomb lattice. In Sec.~\ref{sec:results}, we demonstrate the intriguing tunability of this system by studying orders induced by an {\it in-plane} magnetic field at the hole filling $\nu =2$, namely, the half-filling of the emergent honeycomb lattice. The phase diagram is obtained using the Hartree–Fock analysis of the extended Hubbard model on the emergent honeycomb lattice. Finally, in Sec.~\ref{sec:conclusion}, we summarize our results and experimental signatures of the predicted orders.

\section{Modeling the emergent honeycomb lattice}\label{sec:theory}

\para 
The ABBA-stacked tdbWSe$_2$ can satisfy the three criteria above for the following reasons. First, this structure's increased number of layers leads to stronger interlayer tunnelings compared to a single bilayer, pushing the $\Gamma$-valley to the top of the valence bands~\cite{mak2010atomically,xu2014spin}. Second, inherited from the $D_{3h}$ point group symmetry of each WSe$_2$ layer, the ABBA-stacked tdbWSe$_2$ manifests a three-fold rotation $C_{3z}$ around the $z$-axis, and a two-fold rotation $C_{2y}$ around the in-plane $y$-axis, where the $y$-axis is chosen to align with the $xy$-plane projection of a bond between the tungsten atom and the selenium atom in the monolayer WSe$_2$.
These crystal symmetries ensure a sublattice-symmetric honeycomb lattice composed of the MX and XM sites. Third, from the \textit{ab initio} tight-binding model in Appendix~\ref{app:A}, we find the correct energy hierarchy amongst the MX, XM, and MM sites to ensure an emergent honeycomb lattice.

\para 
To estimate the hopping parameters in the emergent honeycomb lattice based on the three criteria, we need to obtain the dispersion of the two topmost moir\'e valence bands. Adapting the approach in Ref.~\cite{foutty2022tunable,zhang2021electronic}, we begin with a four-layer continuum Hamiltonian for our twisted ABBA-stacked system that preserves $C_{2y}$, $C_{3z}$, {time-reversal} and SU(2) spin rotation symmetries (due to negligible SOC near the $\Gamma$-valley~\cite{fang2015initio}):
\begin{equation}\label{eq:H_4L}
    H_{\text{4L}}=-\frac{\hbar^2 \bm{k}^2}{2m} + \begin{pmatrix}
        \Delta_1(\bm{r}) & \Delta_{12}(\bm{r}) & 0 & 0\\
        \Delta_{12}^\dagger(\bm{r}) & \Delta_2(\bm{r}) & \Delta_{23}(\bm{r}) & 0\\
        0 & \Delta_{23}^\dagger(\bm{r}) & \Delta_3(\bm{r}) & \Delta_{34}(\bm{r}) \\
        0 & 0 & \Delta_{34}^\dagger(\bm{r}) & \Delta_4(\bm{r})
    \end{pmatrix}.
\end{equation} 

\begin{figure}[ht]
    \centering
    \includegraphics[width=3.4in]{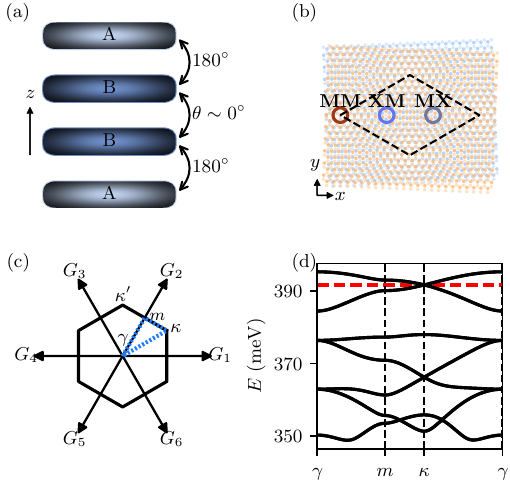}
    \caption{
    (a) The schematic picture of the ABBA-stacked tdbWSe$_2$. Within each of the AB- and BA-stacked bilayer WSe$_2$, the orientation of the two layers 
    differs by 180$^\circ$. The two bilayers are twisted so that the middle two layers have a small twist angle $\theta$ between them.
    (b) The real-space moir\'e pattern created by the twisted central two layers (blue and orange) with the darker (lighter) for the metal (chalcogenide) atom. The three high-symmetry-stacking sites in one moir\'e unit cell (black dashed lines) are MM (brown, metal atoms on one layer aligned with metal atoms on the adjacent layer), XM (blue, chalcogenide atom on one layer aligned with metal atoms on the adjacent layer), and MX (grey, metal atom on one layer aligned with chalcogenide atoms on the adjacent layer). 
    Here, the alignment of atoms refers to the central two layers in the double bilayer structure; 
    (c) Moir\'e Brillouin zone with four high-symmetry points $\gamma$, $m$, $\kappa$, and $\kappa^\prime$. $G_{1,..,6}$ are the moir\'e reciprocal vectors; 
    (d) Moir\'e band structure obtained from the continuum model at twist angle $\theta = 2^\circ$ along the path $\gamma$-$m$-$\kappa$-$\gamma$ [blue dashed line shown in (c)]. The red dashed line indicates Fermi energy at hole filling $\nu=2$ (two holes per mori\'e unit cell). 
    }
    \label{fig:schematic}
\end{figure}

$H_{\text{4L}}$ is identical for both spin species. The intralayer potentials take the forms $\Delta_1(\bm{r})=\Delta_4(\bm{r})=V_{1}$ for the first and fourth layer, and $\Delta_{2}(\bm{r})=V_{2}^{(0)}+2V_{2}^{(1)}\sum\limits_{i=1,3,5}\cos(\bm{G}_i\cdot \bm{r}+\phi)$ and $\Delta_{3}(\bm{r})=V_{2}^{(0)}+2V_{2}^{(1)}\sum\limits_{i=1,3,5}\cos(\bm{G}_i\cdot \bm{r}-\phi)$ for the second and third layer (the opposite sign of $\phi$ as a result of $C_{2y}$ symmetry). 
The interlayer tunnelings take the forms $\Delta_{12}(\bm{r})=\Delta_{34}(\bm{r})=V_{12}$ and $\Delta_{23}(\bm{r})=V_{23}^{(0)}+2 V_{23}^{(1)} \sum\limits_{i=1,3,5}\cos(\bm{G}_i\cdot \bm{r})$. 
Here, $\bm{G}_i=\frac{4\pi}{\sqrt{3}a_M}\left( \cos(\frac{(i-1)\pi}{3}),\sin(\frac{(i-1)\pi}{3}) \right)$ for $i=1,2,...,6$ are the six first-shell moir\'e reciprocal lattice vectors (see Fig. \ref{fig:schematic} (c)). The moir\'e lattice constant $a_M$ is determined by the small twist angle $\theta$ through $a_M \approx a_0/\theta$ with the lattice constant $a_0$ of monolayer WSe$_2$ being 3.28\AA{}~\cite{he2014stacking}. At $\theta\sim 2^\circ$, $a_M \approx 10$ nm. The moir\'e reciprocal lattice vectors $\bm{G}_i$ only show up in $\Delta_2=\Delta_3$ and $\Delta_{23}$ because the small-angle twisting only appears between the second and third layer of WSe$_2$ in the twisted double bilayer structure.

\para 
We now estimate the parameters in the intralayer potentials and interlayer tunnelings in Eq.~\eqref{eq:H_4L}. 
These potentials and tunnelings in the tdbWSe$_2$ evaluated at the MM, and MX/XM sites should be fitted according to the band structures of
the \emph{untwisted} ABBA-stacked double bilayer WSe$_2$ in the three high-symmetry stacking configurations MM, and MX/XM [see Appendix~\ref{app:A} for more details]~\cite{jung2014initio}. 
The angle-independent parameters for the potentials and tunnelings are estimated as follows:
$(V_1,V_2^{(0)},V_2^{(1)})=(200,-159,-8)$ meV, $\phi=-0.17$, and $(V_{12},V_{23}^{(0)},V_{23}^{(1)})=(184, 356, -9)$ meV. We comment that these parameter estimates are obtained based on assumptions on the interlayer distances in the double bilayer WSe$_2$ detailed in App. \ref{app:A}. The assumed interlayer distances produce a representative example of tdbWSe$_2$ satisfying the energetics criteria allowing us to estimate bandwidth and demonstrate the tunability of the emergent honeycomb lattice, which are the main purposes of our work. With the above parameters, we can solve Hamiltonian~\eqref{eq:H_4L} for the moir\'e valence bands of tdbWSe$_2$. For example,  Figure~\ref{fig:schematic}(d) shows the band structure along the $\gamma$-$m$-$\kappa$-$\gamma$ path in the moir\'e Brillouin zone (mBZ) at a twist angle of $\theta = 2^\circ$.
From the two topmost moir\'e valence bands, especially the presence of a Dirac cone between them, we confirm that ABBA-stacked tdbWSe$_2$ with a hole filling less than 4 per moir\'e unit cell captures an emergent honeycomb lattice with both sublattice and SU(2) spin rotation symmetries. At a twist angle of $\theta=2^\circ$, the bandwidth is around 
10 meV (which is much smaller than the $\sim$2 eV bandwidth of graphene~\cite{jung2013tightbinding}).

\begin{figure}[ht]
    \centering
    \includegraphics[width=3.4in]{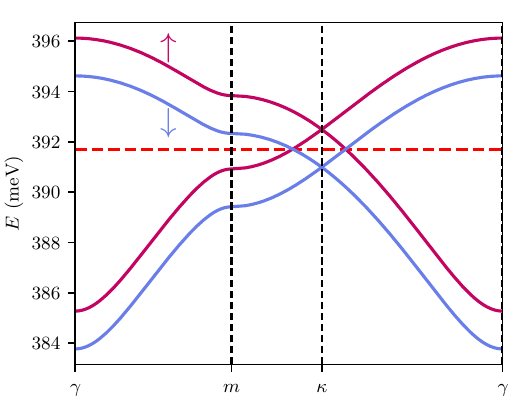}
    \caption{
    The two topmost moir\'e valence bands shown in Fig.~\ref{fig:schematic}(d) 
    split by the Zeeman energy shift due to an in-plane magnetic field of 13 T. The arrows indicate the spin along the in-plane field direction. 
    The red dashed line indicates the Fermi energy at hole filling $\nu=2$, i.e., two holes per moir\'e unit cell and half-filling for the emergent honeycomb lattice. }
    \label{fig:bandstructure}
\end{figure}

\para At hole filling $\nu =2 $, i.e., two holes per moir\'e unit cell, the Fermi level is precisely at the Dirac point protected by $C_{3z}$ and $C_{2y}$ symmetries. This results in a semimetal (non-interacting) ground state, as shown in Fig.~\ref{fig:schematic}(d). The narrow bandwidth of the tdbWSe$_2$-realized honeycomb lattice allows for an intriguing way to tune the electronic structure: a realistic in-plane magnetic field can ``dope" the semimetal and create finite Fermi surfaces located around both $\kappa$ and $\kappa'$ points of the mBZ for both spin species. Note that such tuning is impossible for the $K$/$K'$-valley-based TMD moir\'e system due to the strong SOC therein.
{The ``doping" under the in-plane magnetic field is the result of the Zeeman effect. The thinness of the double-bilayer structure allows us to ignore the orbital effect of the in-plane field.}
Without loss of generality, we assume the in-plane magnetic field $B$ is applied along the $x$ direction, resulting in a Zeeman energy of $ h_x=\frac{1}{2}g \mu_B B$, where $\mu_B$ is the Bohr magneton ($ 5.79 \times10^{-2}$ meV/T,) and the $g$-factor is roughly 2 due to the negligible SOC near the $\Gamma$-valley. 
For example, at a twist angle of $\theta = 2^\circ$, a magnetic field of 13 T, typically achievable experimentally, creates a Zeeman splitting energy of 0.8 meV. This Zeeman splitting leads to a finite Fermi surface of each spin species roughly 
halfway in energy between the Dirac point (at the $\kappa$ point in the mBZ) and the van Hove singularity (at the $m$ point), as shown in Fig.~\ref{fig:bandstructure}. 
In the presence of the in-plane field, a magnetic particle-hole instability that opens a charge gap is expected even at weak interactions because of the nesting between the Fermi surfaces of the two spin species. This instability would lead to a magnetic order that spontaneously breaks the remaining U(1) spin rotation symmetry about the $x$-axis, namely, the in-plane field direction.

\para Conceptually, a similar in-plane-field-induced particle-hole instability applies to graphene. The previous work \cite{bercx2009magnetic} shows that the resulting state is a canted N\'eel antiferromagnetic (AF) insulator, whose magnetizations projected onto the plane perpendicular to the applied field are opposite in the two sublattices of the honeycomb. However, realizing such a state in graphene requires a very high in-plane field of $10^2\sim 10^3$ T \cite{bercx2009magnetic}. As we show below, a similar canted N\'eel AF insulator can be found in the tdbWSe$_2$-realized honeycomb lattice, whose bandwidth is engineered around the meV scale, at a realistic in-plane field of a few Tesla. 

\section{Extended Hubbard model and phase diagram under in-plane field}\label{sec:results}

\para 
Aside from the bandwidth distinction with graphene, the band structure of the tdbWSe$_2$-realized honeycomb lattice exhibits a more pronounced contribution from the next-nearest-neighbor (NNN) hopping $t_2$ on the honeycomb lattice. The effect of $t_2$ is manifested by the asymmetry (about the Dirac cone) between the two topmost valence bands shown in Fig. \ref{fig:schematic} (d). In contrast to the graphene-inspired model of \cite{bercx2009magnetic} with only the nearest-neighbor (NN) hopping $t_1$ and the on-site Hubbard interaction, we consider an extended Hubbard model with longer-range hoppings and interactions pertaining to our tdbWSe$_2$-realized honeycomb lattice system. Our Hamiltonian consists of three terms: the hopping $H_{\rm h}$, the Zeeman energy $H_{\rm Z}$ from the in-plane magnetic field, and the extended Hubbard interaction $H_{\text{int}}$. The term $H_{\rm h} = -t_1 \sum_{\expval{ij}}\sum_{\sigma} c_{i\sigma}^\dagger c_{j\sigma}  - t_2 \sum_{\expval{\expval{ij}}}\sum_{\sigma} c_{i\sigma}^\dagger c_{j\sigma} $ contains both NN hopping $t_1$ and NNN hoppings $t_2$. Here, $c_{i\sigma}$ is the fermion operator of the hole with spin $\sigma$ at the site $i$ of the emergent honeycomb lattice. {Both $t_1$ and $t_2$ are real because of the SU(2) spin rotation symmetry and the time-reversal symmetry. $t_{1,2}$ are}
are tunable by the twist angle $\theta$.
For instance,  at $\theta=2^\circ$, 
we estimate $t_1=1.8$ meV and $t_2=0.2$ meV based on the moir\'e band structure obtained from the continuum model Eq.\eqref{eq:H_4L}. The ratio $t_2/t_1$ varies as we change the twist angle: 0.018 for $\theta=1^\circ$, 0.11 for $\theta=2^\circ$, and 0.12 for $\theta=3^\circ$. In the search for possible phases below, we treat $t_2/t_1$ as a parameter to explore the effect of longer-range hopping. The Zeeman energy $H_{\rm Z}$ is given by $H_{\rm Z}=h_x  \sum_{i} \left( c_{i\uparrow}^\dagger c_{i\uparrow} - c_{i\downarrow}^\dagger c_{i\downarrow} \right) $, where the spin quantization axis conveniently is chosen along the $x$ direction. Under $H_{\rm Z}$, the spin symmetry is reduced to a U(1) spin rotation in the $yz$ plane. We fix $h_x=0.4 t_1$ in the following. For the twist angle $\theta=2^\circ$, $h_x=0.4 t_1$ (corresponding to a 13 T in-plane magnetic field) 
sets the Fermi level roughly halfway between the Dirac point at the $\kappa/\kappa'$ point and the van Hove singularity at the $m$ point of the mBZ.

\para The range of interactions in TMD-based superlattices typically extends beyond on-site interactions. Here, we consider the extended Hubbard interaction  
\begin{equation}\label{eq:Hint}
    H_{\text{int}}=U_0 \sum_i c_{i\uparrow}^\dagger c_{i\uparrow} c_{i\downarrow}^\dagger c_{i\downarrow} + U_1 \sum_{\langle ij\rangle}\sum_{\sigma,\sigma'=\uparrow,\downarrow} c_{i,\sigma}^\dagger c_{i,\sigma} c_{j,\sigma'}^\dagger c_{j,\sigma'} ,
\end{equation}
with the on-site interaction strength $U_0$, and the NN Hubbard interaction of strength $U_1$. In the following study of the phase diagram, we treat $U_0$ and $U_1$ as two free parameters to explore their effect on the canted N\'eel AF insulator and other phases.

\begin{figure*}[ht]
    \centering
     \includegraphics[width=6.8in]{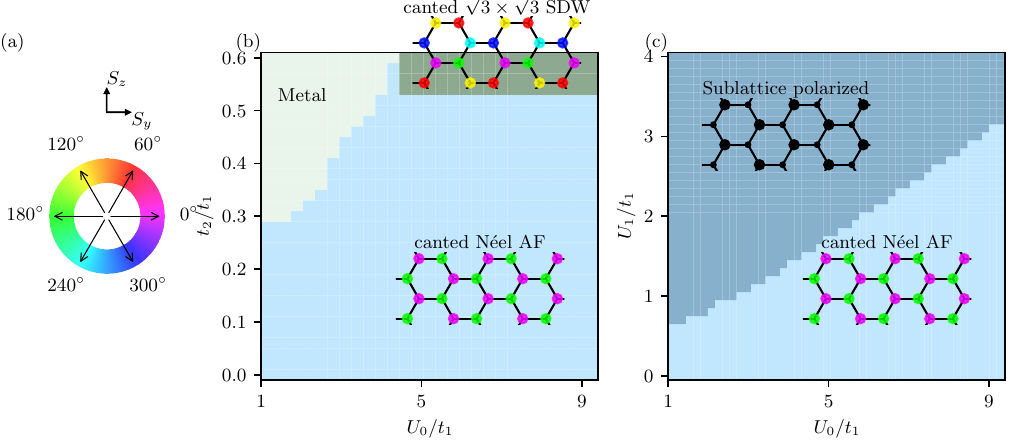}
    \caption{
        (a) The color wheel shows the six spin orientations in the $yz$ plane.
        (b) The phase diagram as a function of $U_0$ and $t_2$. (c) The phase diagram as a function of $U_0$ and $U_1$. Here, the size of the dot indicates the number density of each site, while the color indicates the spin orientation in the $yz$ plane as shown in the color wheel. The black dots in the sublattice polarized state indicate zero spin polarization.}
    \label{fig:phasediagram}
\end{figure*}

\para We start with understanding the effect of the NNN hopping $t_2$, and present the ground-state phase diagram obtained using the Hartree-Fock method in Fig.~\ref{fig:phasediagram}(b). In this phase diagram,
we consider a parameter range $U_0 \in [t_1, 9t_1]$ and $t_2 \in [0, 0.6t_1]$ while setting $U_1 = 0$. Using the numerical Hartree-Fock method, we find three possible phases: the canted N\'eel AF insulator, canted $\sqrt{3}\times\sqrt{3}$ spin density wave (SDW), and a metal. At hole filling $\nu = 2$, the Fermi surfaces of the two spin species are always perfectly nested for $t_2 \lesssim t_1/3$ at both the $\kappa$ and $\kappa'$ points of the mBZ. This perfect nesting results in a Fermi surface instability towards a robust canted N\'eel AF insulator, spontaneously breaking the U(1) spin rotation symmetry in the $yz$ plane as mentioned above. 
Beyond $t_2>t_1/3$, the system's ground state is metallic at small $U_0$. The metallic state is due to an extra Fermi surface of the spin-down hole appearing around the $\gamma$ point when $t_2>t_1/3$. Consequently, the Fermi surfaces of the two spin species are no longer nested at $\nu =2$. As $U_0$ increases, the system can enter an insulating SDW phase with an enlarged $\sqrt{3} \times \sqrt{3}$ unit cell. This SDW phase exhibits a 120$^\circ$ AF order in the $yz$ plane on each sublattice of the honeycomb. The chiralities of the two 120$^\circ$ AF orders are opposite. For the tdbWSe$_2$-realized honeycomb lattice, $t_2/t_1<1/3$ for twist angle $\theta<4^\circ$. Hence, restricted to only on-site interactions, the canted N\'eel AF insulator is the most relevant in-plane-field-induced phase for this material at hole filling $\nu=2$.

\para In addition to the NNN hopping $t_2$, we also study the effect of the NN Hubbard interaction $U_1$. We present the phase diagram obtained from the Hartree-Fock calculation as a function of $U_0$ and $U_1$ as shown in Fig.~\ref{fig:phasediagram}(c). For this phase diagram, we set $t_2 = 0$ for simplicity. The canted N\'eel AF insulator is robust at small $U_1$. As $U_1$ increases, the system is driven to an insulating phase with the charge distribution polarized to one of the sublattices, spontaneously breaking the system's $C_{2y}$ symmetry. In contrast to the canted N\'eel AF insulator, the $yz$-plane U(1) spin rotation symmetry is preserved in this phase. The phase boundary between the AF insulator and the sublattice-polarized insulator is roughly at $U_1\sim \frac{1}{3}U_0 + \frac{2}{3}h_x$.
This is the exact phase boundary in the strong interaction limit (where the hopping terms $H_{\rm h}$ are neglected). In the tdbWSe$_2$-realized honeycomb
lattice, the screening of the Coulomb interaction determines the relative strength between $U_0$ and $U_1$. Stronger screening favors the canted N\'eel AF insulator, while weaker screening favors the sublattice-polarized insulator. In Appendix~\ref{app:B}, we present an analytical mean-field analysis in the weak-coupling limit $U_{0,1} \ll t_1$ and find that the canted $\sqrt{3}\times\sqrt{3}$ SDW insulator can also compete with the canted N\'eel AF insulator when $U_1 /U_0 $ increases.

\section{Summary and Experimental Implication}\label{sec:conclusion}
\para In summary, we show that the ABBA-stacked tdbWSe$_2$ can serve as a realistic and tunable platform to simulate $\Gamma$-valley honeycomb lattice with both sublattice and SU(2) spin rotation symmetries. We develop a continuum model to estimate the relevant hopping parameters of this honeycomb lattice model, including the NN and the NNN hoppings $t_{1,2}$. The small bandwidth, found to be at the meV scale, and the ratio $t_2/t_1$ are both tunable via the twist angle $\theta$. We show that the small bandwidth enables the effective control of the electronic structure via an in-plane magnetic field. At a hole filling $\nu=2$, an in-plane field of a few Teslas can dope this system from a Dirac semimetal to a metal having finite Fermi surfaces with instabilities. To demonstrate this tunability and understand the resulting instabilities, we construct an extended Hubbard model for this honeycomb lattice and perform a numerical Hartree-Fock calculation for the ground-state phase diagram at $\nu=2$. We find a robust canted N\'eel AF insulating phase when the on-site Hubbard interaction $U_0$ dominates. This phase is conceptually similar to the magnetic-field-induced canted AF insulator studied in the context of graphene by Ref. \cite{bercx2009magnetic}. However, our finding resides in a more realistic parameter regime. Moreover, a competing sublattice-polarized insulator is found in our extended Hubbard model when the NN Hubbard interaction $U_1$ increases beyond a threshold.

\para 
The results presented in this work have direct experimental implications. First, for ABBA-stacked tdbWSe$_2$ at hole filling $\nu=2$, an in-plane magnetic field is predicted to induce an insulating ground state. There are two possible insulating ground states: the canted N\'eel AF insulator and the sublattice-polarized insulator. The former can be detected through its spin texture perpendicular to the in-plane field direction, potentially using spin-polarized scanning tunneling microscopy~\cite{pietzsch2001observation}. The finite-temperature phase transition into the canted N\'eel AF order is a Berezinskii–Kosterlitz–Thouless transition associated with the remaining U(1) spin rotation symmetry under the in-plane field. 
This metal-insulator transition can be accessed by lowering the temperature at a given in-plane field or by increasing the in-plane field at a low temperature starting from the Dirac semi-metal.
{To be more precise, at zero temperature (and in the absence of disorder), there should be no threshold for the interaction strength or in-plane magnetic field --- as long as they are both finite, the insulator phase exists. However, at a finite temperature, one has to turn on an in-plane field and interaction strength beyond certain thresholds to observe the insulating phase with the spin orders. The thresholds can be estimated by comparing the temperature with the gap sizes of the insulating phase at zero temperature. Therefore, at sufficiently low-temperature, one can access a metal-insulator transition using an in-plane magnetic field.}
The sublattice-polarized insulator has no non-trivial spin texture (other than the magnetization along the in-plane field direction). Instead, the spontaneous charge polarization in the two sublattices of the emergent honeycomb lattice leads to a finite electric polarization in the tdbWSe$_2$ sample along the $z$ direction. Consequently, a signature of this sublattice-polarized insulating phase is the hysteresis of 
the $z$-axis electric polarization as a function of the $z$-direcction electric field. The finite-temperature phase transition into the sublattice-polarized phase is in the Ising universality class. Switching between the two insulating phases can be implemented by tuning the ratio between $U_0$ and $U_1$, which is achievable via adjusting the twist angle and/or sample's distance from the gates.

{\bf Acknowledgements.}
We thank Kin Fai Mak, Jie Shan, and Liguo Ma for stimulating discussions. H.P. acknowledges the helpful discussions with Fengcheng Wu, Ming Xie, and Dan Mao. 
H.P. and E-A.K. acknowledge the funding support National Science Foundation (Platform for the Accelerated Realization, Analysis, and Discovery of Interface Materials (PARADIM)) under Cooperative Agreement No. DMR-1539918. C.-M.J. is supported by a faculty startup grant at Cornell University.

\bibliography{ABBA.bib}

\begin{thebibliography}{38}%
\makeatletter
\providecommand \@ifxundefined [1]{%
 \@ifx{#1\undefined}
}%
\providecommand \@ifnum [1]{%
 \ifnum #1\expandafter \@firstoftwo
 \else \expandafter \@secondoftwo
 \fi
}%
\providecommand \@ifx [1]{%
 \ifx #1\expandafter \@firstoftwo
 \else \expandafter \@secondoftwo
 \fi
}%
\providecommand \natexlab [1]{#1}%
\providecommand \enquote  [1]{``#1''}%
\providecommand \bibnamefont  [1]{#1}%
\providecommand \bibfnamefont [1]{#1}%
\providecommand \citenamefont [1]{#1}%
\providecommand \href@noop [0]{\@secondoftwo}%
\providecommand \href [0]{\begingroup \@sanitize@url \@href}%
\providecommand \@href[1]{\@@startlink{#1}\@@href}%
\providecommand \@@href[1]{\endgroup#1\@@endlink}%
\providecommand \@sanitize@url [0]{\catcode `\\12\catcode `\$12\catcode
  `\&12\catcode `\#12\catcode `\^12\catcode `\_12\catcode `\%12\relax}%
\providecommand \@@startlink[1]{}%
\providecommand \@@endlink[0]{}%
\providecommand \url  [0]{\begingroup\@sanitize@url \@url }%
\providecommand \@url [1]{\endgroup\@href {#1}{\urlprefix }}%
\providecommand \urlprefix  [0]{URL }%
\providecommand \Eprint [0]{\href }%
\providecommand \doibase [0]{https://doi.org/}%
\providecommand \selectlanguage [0]{\@gobble}%
\providecommand \bibinfo  [0]{\@secondoftwo}%
\providecommand \bibfield  [0]{\@secondoftwo}%
\providecommand \translation [1]{[#1]}%
\providecommand \BibitemOpen [0]{}%
\providecommand \bibitemStop [0]{}%
\providecommand \bibitemNoStop [0]{.\EOS\space}%
\providecommand \EOS [0]{\spacefactor3000\relax}%
\providecommand \BibitemShut  [1]{\csname bibitem#1\endcsname}%
\let\auto@bib@innerbib\@empty
\bibitem [{\citenamefont {Honerkamp}(2008)}]{honerkamp2008density}%
  \BibitemOpen
  \bibfield  {author} {\bibinfo {author} {\bibfnamefont {C.}~\bibnamefont
  {Honerkamp}},\ }\bibfield  {title} {\bibinfo {title} {Density {{Waves}} and
  {{Cooper Pairing}} on the {{Honeycomb Lattice}}},\ }\href
  {https://doi.org/10.1103/PhysRevLett.100.146404} {\bibfield  {journal}
  {\bibinfo  {journal} {Phys. Rev. Lett.}\ }\textbf {\bibinfo {volume} {100}},\
  \bibinfo {pages} {146404} (\bibinfo {year} {2008})}\BibitemShut {NoStop}%
\bibitem [{\citenamefont {Aleiner}\ \emph {et~al.}(2007)\citenamefont
  {Aleiner}, \citenamefont {Kharzeev},\ and\ \citenamefont
  {Tsvelik}}]{aleiner2007spontaneous}%
  \BibitemOpen
  \bibfield  {author} {\bibinfo {author} {\bibfnamefont {I.~L.}\ \bibnamefont
  {Aleiner}}, \bibinfo {author} {\bibfnamefont {D.~E.}\ \bibnamefont
  {Kharzeev}},\ and\ \bibinfo {author} {\bibfnamefont {A.~M.}\ \bibnamefont
  {Tsvelik}},\ }\bibfield  {title} {\bibinfo {title} {Spontaneous symmetry
  breaking in graphene subjected to an in-plane magnetic field},\ }\href
  {https://doi.org/10.1103/PhysRevB.76.195415} {\bibfield  {journal} {\bibinfo
  {journal} {Phys. Rev. B}\ }\textbf {\bibinfo {volume} {76}},\ \bibinfo
  {pages} {195415} (\bibinfo {year} {2007})}\BibitemShut {NoStop}%
\bibitem [{\citenamefont {Bercx}\ \emph {et~al.}(2009)\citenamefont {Bercx},
  \citenamefont {Lang},\ and\ \citenamefont {Assaad}}]{bercx2009magnetic}%
  \BibitemOpen
  \bibfield  {author} {\bibinfo {author} {\bibfnamefont {M.}~\bibnamefont
  {Bercx}}, \bibinfo {author} {\bibfnamefont {T.~C.}\ \bibnamefont {Lang}},\
  and\ \bibinfo {author} {\bibfnamefont {F.~F.}\ \bibnamefont {Assaad}},\
  }\bibfield  {title} {\bibinfo {title} {Magnetic field induced
  semimetal-to-canted-antiferromagnet transition on the honeycomb lattice},\
  }\href {https://doi.org/10.1103/PhysRevB.80.045412} {\bibfield  {journal}
  {\bibinfo  {journal} {Phys. Rev. B}\ }\textbf {\bibinfo {volume} {80}},\
  \bibinfo {pages} {045412} (\bibinfo {year} {2009})}\BibitemShut {NoStop}%
\bibitem [{\citenamefont {Nandkishore}\ \emph {et~al.}(2012)\citenamefont
  {Nandkishore}, \citenamefont {Levitov},\ and\ \citenamefont
  {Chubukov}}]{nandkishore2012chiral}%
  \BibitemOpen
  \bibfield  {author} {\bibinfo {author} {\bibfnamefont {R.}~\bibnamefont
  {Nandkishore}}, \bibinfo {author} {\bibfnamefont {L.~S.}\ \bibnamefont
  {Levitov}},\ and\ \bibinfo {author} {\bibfnamefont {A.~V.}\ \bibnamefont
  {Chubukov}},\ }\bibfield  {title} {\bibinfo {title} {Chiral superconductivity
  from repulsive interactions in doped graphene},\ }\href
  {https://doi.org/10.1038/nphys2208} {\bibfield  {journal} {\bibinfo
  {journal} {Nature Physics}\ }\textbf {\bibinfo {volume} {8}},\ \bibinfo
  {pages} {158} (\bibinfo {year} {2012})}\BibitemShut {NoStop}%
\bibitem [{\citenamefont {Nandkishore}\ \emph {et~al.}(2014)\citenamefont
  {Nandkishore}, \citenamefont {Thomale},\ and\ \citenamefont
  {Chubukov}}]{nandkishore2014superconductivity}%
  \BibitemOpen
  \bibfield  {author} {\bibinfo {author} {\bibfnamefont {R.}~\bibnamefont
  {Nandkishore}}, \bibinfo {author} {\bibfnamefont {R.}~\bibnamefont
  {Thomale}},\ and\ \bibinfo {author} {\bibfnamefont {A.~V.}\ \bibnamefont
  {Chubukov}},\ }\bibfield  {title} {\bibinfo {title} {Superconductivity from
  weak repulsion in hexagonal lattice systems},\ }\href
  {https://doi.org/10.1103/PhysRevB.89.144501} {\bibfield  {journal} {\bibinfo
  {journal} {Phys. Rev. B}\ }\textbf {\bibinfo {volume} {89}},\ \bibinfo
  {pages} {144501} (\bibinfo {year} {2014})}\BibitemShut {NoStop}%
\bibitem [{\citenamefont {Raghu}\ \emph {et~al.}(2008)\citenamefont {Raghu},
  \citenamefont {Qi}, \citenamefont {Honerkamp},\ and\ \citenamefont
  {Zhang}}]{raghu2008topological}%
  \BibitemOpen
  \bibfield  {author} {\bibinfo {author} {\bibfnamefont {S.}~\bibnamefont
  {Raghu}}, \bibinfo {author} {\bibfnamefont {X.-L.}\ \bibnamefont {Qi}},
  \bibinfo {author} {\bibfnamefont {C.}~\bibnamefont {Honerkamp}},\ and\
  \bibinfo {author} {\bibfnamefont {S.-C.}\ \bibnamefont {Zhang}},\ }\bibfield
  {title} {\bibinfo {title} {Topological {{Mott Insulators}}},\ }\href
  {https://doi.org/10.1103/PhysRevLett.100.156401} {\bibfield  {journal}
  {\bibinfo  {journal} {Phys. Rev. Lett.}\ }\textbf {\bibinfo {volume} {100}},\
  \bibinfo {pages} {156401} (\bibinfo {year} {2008})}\BibitemShut {NoStop}%
\bibitem [{\citenamefont {Regan}\ \emph {et~al.}(2020)\citenamefont {Regan},
  \citenamefont {Wang}, \citenamefont {Jin}, \citenamefont {Bakti~Utama},
  \citenamefont {Gao}, \citenamefont {Wei}, \citenamefont {Zhao}, \citenamefont
  {Zhao}, \citenamefont {Zhang}, \citenamefont {Yumigeta}, \citenamefont
  {Blei}, \citenamefont {Carlstr{\"o}m}, \citenamefont {Watanabe},
  \citenamefont {Taniguchi}, \citenamefont {Tongay}, \citenamefont {Crommie},
  \citenamefont {Zettl},\ and\ \citenamefont {Wang}}]{regan2020mott}%
  \BibitemOpen
  \bibfield  {author} {\bibinfo {author} {\bibfnamefont {E.~C.}\ \bibnamefont
  {Regan}}, \bibinfo {author} {\bibfnamefont {D.}~\bibnamefont {Wang}},
  \bibinfo {author} {\bibfnamefont {C.}~\bibnamefont {Jin}}, \bibinfo {author}
  {\bibfnamefont {M.~I.}\ \bibnamefont {Bakti~Utama}}, \bibinfo {author}
  {\bibfnamefont {B.}~\bibnamefont {Gao}}, \bibinfo {author} {\bibfnamefont
  {X.}~\bibnamefont {Wei}}, \bibinfo {author} {\bibfnamefont {S.}~\bibnamefont
  {Zhao}}, \bibinfo {author} {\bibfnamefont {W.}~\bibnamefont {Zhao}}, \bibinfo
  {author} {\bibfnamefont {Z.}~\bibnamefont {Zhang}}, \bibinfo {author}
  {\bibfnamefont {K.}~\bibnamefont {Yumigeta}}, \bibinfo {author}
  {\bibfnamefont {M.}~\bibnamefont {Blei}}, \bibinfo {author} {\bibfnamefont
  {J.~D.}\ \bibnamefont {Carlstr{\"o}m}}, \bibinfo {author} {\bibfnamefont
  {K.}~\bibnamefont {Watanabe}}, \bibinfo {author} {\bibfnamefont
  {T.}~\bibnamefont {Taniguchi}}, \bibinfo {author} {\bibfnamefont
  {S.}~\bibnamefont {Tongay}}, \bibinfo {author} {\bibfnamefont
  {M.}~\bibnamefont {Crommie}}, \bibinfo {author} {\bibfnamefont
  {A.}~\bibnamefont {Zettl}},\ and\ \bibinfo {author} {\bibfnamefont
  {F.}~\bibnamefont {Wang}},\ }\bibfield  {title} {\bibinfo {title} {Mott and
  generalized {{Wigner}} crystal states in {{WSe}}{$_2$}/{{WS}}{$_2$} moir\'e
  superlattices},\ }\href {https://doi.org/10.1038/s41586-020-2092-4}
  {\bibfield  {journal} {\bibinfo  {journal} {Nature}\ }\textbf {\bibinfo
  {volume} {579}},\ \bibinfo {pages} {359} (\bibinfo {year}
  {2020})}\BibitemShut {NoStop}%
\bibitem [{\citenamefont {Tang}\ \emph {et~al.}(2020)\citenamefont {Tang},
  \citenamefont {Li}, \citenamefont {Li}, \citenamefont {Xu}, \citenamefont
  {Liu}, \citenamefont {Barmak}, \citenamefont {Watanabe}, \citenamefont
  {Taniguchi}, \citenamefont {MacDonald}, \citenamefont {Shan},\ and\
  \citenamefont {Mak}}]{tang2020simulation}%
  \BibitemOpen
  \bibfield  {author} {\bibinfo {author} {\bibfnamefont {Y.}~\bibnamefont
  {Tang}}, \bibinfo {author} {\bibfnamefont {L.}~\bibnamefont {Li}}, \bibinfo
  {author} {\bibfnamefont {T.}~\bibnamefont {Li}}, \bibinfo {author}
  {\bibfnamefont {Y.}~\bibnamefont {Xu}}, \bibinfo {author} {\bibfnamefont
  {S.}~\bibnamefont {Liu}}, \bibinfo {author} {\bibfnamefont {K.}~\bibnamefont
  {Barmak}}, \bibinfo {author} {\bibfnamefont {K.}~\bibnamefont {Watanabe}},
  \bibinfo {author} {\bibfnamefont {T.}~\bibnamefont {Taniguchi}}, \bibinfo
  {author} {\bibfnamefont {A.~H.}\ \bibnamefont {MacDonald}}, \bibinfo {author}
  {\bibfnamefont {J.}~\bibnamefont {Shan}},\ and\ \bibinfo {author}
  {\bibfnamefont {K.~F.}\ \bibnamefont {Mak}},\ }\bibfield  {title} {\bibinfo
  {title} {Simulation of {{Hubbard}} model physics in
  {{WSe}}{$_2$}/{{WS}}{$_2$} moir\'e superlattices},\ }\href
  {https://doi.org/10.1038/s41586-020-2085-3} {\bibfield  {journal} {\bibinfo
  {journal} {Nature}\ }\textbf {\bibinfo {volume} {579}},\ \bibinfo {pages}
  {353} (\bibinfo {year} {2020})}\BibitemShut {NoStop}%
\bibitem [{\citenamefont {Xu}\ \emph {et~al.}(2020)\citenamefont {Xu},
  \citenamefont {Liu}, \citenamefont {Rhodes}, \citenamefont {Watanabe},
  \citenamefont {Taniguchi}, \citenamefont {Hone}, \citenamefont {Elser},
  \citenamefont {Mak},\ and\ \citenamefont {Shan}}]{xu2020correlated}%
  \BibitemOpen
  \bibfield  {author} {\bibinfo {author} {\bibfnamefont {Y.}~\bibnamefont
  {Xu}}, \bibinfo {author} {\bibfnamefont {S.}~\bibnamefont {Liu}}, \bibinfo
  {author} {\bibfnamefont {D.~A.}\ \bibnamefont {Rhodes}}, \bibinfo {author}
  {\bibfnamefont {K.}~\bibnamefont {Watanabe}}, \bibinfo {author}
  {\bibfnamefont {T.}~\bibnamefont {Taniguchi}}, \bibinfo {author}
  {\bibfnamefont {J.}~\bibnamefont {Hone}}, \bibinfo {author} {\bibfnamefont
  {V.}~\bibnamefont {Elser}}, \bibinfo {author} {\bibfnamefont {K.~F.}\
  \bibnamefont {Mak}},\ and\ \bibinfo {author} {\bibfnamefont {J.}~\bibnamefont
  {Shan}},\ }\bibfield  {title} {\bibinfo {title} {Correlated insulating states
  at fractional fillings of moir\'e superlattices},\ }\href
  {https://doi.org/10.1038/s41586-020-2868-6} {\bibfield  {journal} {\bibinfo
  {journal} {Nature}\ }\textbf {\bibinfo {volume} {587}},\ \bibinfo {pages}
  {214} (\bibinfo {year} {2020})}\BibitemShut {NoStop}%
\bibitem [{\citenamefont {Wang}\ \emph {et~al.}(2020)\citenamefont {Wang},
  \citenamefont {Shih}, \citenamefont {Ghiotto}, \citenamefont {Xian},
  \citenamefont {Rhodes}, \citenamefont {Tan}, \citenamefont {Claassen},
  \citenamefont {Kennes}, \citenamefont {Bai}, \citenamefont {Kim},
  \citenamefont {Watanabe}, \citenamefont {Taniguchi}, \citenamefont {Zhu},
  \citenamefont {Hone}, \citenamefont {Rubio}, \citenamefont {Pasupathy},\ and\
  \citenamefont {Dean}}]{wang2020correlated}%
  \BibitemOpen
  \bibfield  {author} {\bibinfo {author} {\bibfnamefont {L.}~\bibnamefont
  {Wang}}, \bibinfo {author} {\bibfnamefont {E.-M.}\ \bibnamefont {Shih}},
  \bibinfo {author} {\bibfnamefont {A.}~\bibnamefont {Ghiotto}}, \bibinfo
  {author} {\bibfnamefont {L.}~\bibnamefont {Xian}}, \bibinfo {author}
  {\bibfnamefont {D.~A.}\ \bibnamefont {Rhodes}}, \bibinfo {author}
  {\bibfnamefont {C.}~\bibnamefont {Tan}}, \bibinfo {author} {\bibfnamefont
  {M.}~\bibnamefont {Claassen}}, \bibinfo {author} {\bibfnamefont {D.~M.}\
  \bibnamefont {Kennes}}, \bibinfo {author} {\bibfnamefont {Y.}~\bibnamefont
  {Bai}}, \bibinfo {author} {\bibfnamefont {B.}~\bibnamefont {Kim}}, \bibinfo
  {author} {\bibfnamefont {K.}~\bibnamefont {Watanabe}}, \bibinfo {author}
  {\bibfnamefont {T.}~\bibnamefont {Taniguchi}}, \bibinfo {author}
  {\bibfnamefont {X.}~\bibnamefont {Zhu}}, \bibinfo {author} {\bibfnamefont
  {J.}~\bibnamefont {Hone}}, \bibinfo {author} {\bibfnamefont {A.}~\bibnamefont
  {Rubio}}, \bibinfo {author} {\bibfnamefont {A.~N.}\ \bibnamefont
  {Pasupathy}},\ and\ \bibinfo {author} {\bibfnamefont {C.~R.}\ \bibnamefont
  {Dean}},\ }\bibfield  {title} {\bibinfo {title} {Correlated electronic phases
  in twisted bilayer transition metal dichalcogenides},\ }\href
  {https://doi.org/10.1038/s41563-020-0708-6} {\bibfield  {journal} {\bibinfo
  {journal} {Nature Materials}\ }\textbf {\bibinfo {volume} {19}},\ \bibinfo
  {pages} {861} (\bibinfo {year} {2020})}\BibitemShut {NoStop}%
\bibitem [{\citenamefont {Ghiotto}\ \emph {et~al.}(2021)\citenamefont
  {Ghiotto}, \citenamefont {Shih}, \citenamefont {Pereira}, \citenamefont
  {Rhodes}, \citenamefont {Kim}, \citenamefont {Zang}, \citenamefont {Millis},
  \citenamefont {Watanabe}, \citenamefont {Taniguchi}, \citenamefont {Hone},
  \citenamefont {Wang}, \citenamefont {Dean},\ and\ \citenamefont
  {Pasupathy}}]{ghiotto2021quantum}%
  \BibitemOpen
  \bibfield  {author} {\bibinfo {author} {\bibfnamefont {A.}~\bibnamefont
  {Ghiotto}}, \bibinfo {author} {\bibfnamefont {E.-M.}\ \bibnamefont {Shih}},
  \bibinfo {author} {\bibfnamefont {G.~S. S.~G.}\ \bibnamefont {Pereira}},
  \bibinfo {author} {\bibfnamefont {D.~A.}\ \bibnamefont {Rhodes}}, \bibinfo
  {author} {\bibfnamefont {B.}~\bibnamefont {Kim}}, \bibinfo {author}
  {\bibfnamefont {J.}~\bibnamefont {Zang}}, \bibinfo {author} {\bibfnamefont
  {A.~J.}\ \bibnamefont {Millis}}, \bibinfo {author} {\bibfnamefont
  {K.}~\bibnamefont {Watanabe}}, \bibinfo {author} {\bibfnamefont
  {T.}~\bibnamefont {Taniguchi}}, \bibinfo {author} {\bibfnamefont {J.~C.}\
  \bibnamefont {Hone}}, \bibinfo {author} {\bibfnamefont {L.}~\bibnamefont
  {Wang}}, \bibinfo {author} {\bibfnamefont {C.~R.}\ \bibnamefont {Dean}},\
  and\ \bibinfo {author} {\bibfnamefont {A.~N.}\ \bibnamefont {Pasupathy}},\
  }\bibfield  {title} {\bibinfo {title} {Quantum criticality in twisted
  transition metal dichalcogenides},\ }\href
  {https://doi.org/10.1038/s41586-021-03815-6} {\bibfield  {journal} {\bibinfo
  {journal} {Nature}\ }\textbf {\bibinfo {volume} {597}},\ \bibinfo {pages}
  {345} (\bibinfo {year} {2021})}\BibitemShut {NoStop}%
\bibitem [{\citenamefont {Zhou}\ \emph {et~al.}(2021)\citenamefont {Zhou},
  \citenamefont {Sung}, \citenamefont {Brutschea}, \citenamefont {Esterlis},
  \citenamefont {Wang}, \citenamefont {Scuri}, \citenamefont {Gelly},
  \citenamefont {Heo}, \citenamefont {Taniguchi}, \citenamefont {Watanabe},
  \citenamefont {Zar{\'a}nd}, \citenamefont {Lukin}, \citenamefont {Kim},
  \citenamefont {Demler},\ and\ \citenamefont {Park}}]{zhou2021bilayer}%
  \BibitemOpen
  \bibfield  {author} {\bibinfo {author} {\bibfnamefont {Y.}~\bibnamefont
  {Zhou}}, \bibinfo {author} {\bibfnamefont {J.}~\bibnamefont {Sung}}, \bibinfo
  {author} {\bibfnamefont {E.}~\bibnamefont {Brutschea}}, \bibinfo {author}
  {\bibfnamefont {I.}~\bibnamefont {Esterlis}}, \bibinfo {author}
  {\bibfnamefont {Y.}~\bibnamefont {Wang}}, \bibinfo {author} {\bibfnamefont
  {G.}~\bibnamefont {Scuri}}, \bibinfo {author} {\bibfnamefont {R.~J.}\
  \bibnamefont {Gelly}}, \bibinfo {author} {\bibfnamefont {H.}~\bibnamefont
  {Heo}}, \bibinfo {author} {\bibfnamefont {T.}~\bibnamefont {Taniguchi}},
  \bibinfo {author} {\bibfnamefont {K.}~\bibnamefont {Watanabe}}, \bibinfo
  {author} {\bibfnamefont {G.}~\bibnamefont {Zar{\'a}nd}}, \bibinfo {author}
  {\bibfnamefont {M.~D.}\ \bibnamefont {Lukin}}, \bibinfo {author}
  {\bibfnamefont {P.}~\bibnamefont {Kim}}, \bibinfo {author} {\bibfnamefont
  {E.}~\bibnamefont {Demler}},\ and\ \bibinfo {author} {\bibfnamefont
  {H.}~\bibnamefont {Park}},\ }\bibfield  {title} {\bibinfo {title} {Bilayer
  {{Wigner}} crystals in a transition metal dichalcogenide heterostructure},\
  }\href {https://doi.org/10.1038/s41586-021-03560-w} {\bibfield  {journal}
  {\bibinfo  {journal} {Nature}\ }\textbf {\bibinfo {volume} {595}},\ \bibinfo
  {pages} {48} (\bibinfo {year} {2021})}\BibitemShut {NoStop}%
\bibitem [{\citenamefont {Li}\ \emph {et~al.}(2021{\natexlab{a}})\citenamefont
  {Li}, \citenamefont {Li}, \citenamefont {Regan}, \citenamefont {Wang},
  \citenamefont {Zhao}, \citenamefont {Kahn}, \citenamefont {Yumigeta},
  \citenamefont {Blei}, \citenamefont {Taniguchi}, \citenamefont {Watanabe},
  \citenamefont {Tongay}, \citenamefont {Zettl}, \citenamefont {Crommie},\ and\
  \citenamefont {Wang}}]{li2021imaging}%
  \BibitemOpen
  \bibfield  {author} {\bibinfo {author} {\bibfnamefont {H.}~\bibnamefont
  {Li}}, \bibinfo {author} {\bibfnamefont {S.}~\bibnamefont {Li}}, \bibinfo
  {author} {\bibfnamefont {E.~C.}\ \bibnamefont {Regan}}, \bibinfo {author}
  {\bibfnamefont {D.}~\bibnamefont {Wang}}, \bibinfo {author} {\bibfnamefont
  {W.}~\bibnamefont {Zhao}}, \bibinfo {author} {\bibfnamefont {S.}~\bibnamefont
  {Kahn}}, \bibinfo {author} {\bibfnamefont {K.}~\bibnamefont {Yumigeta}},
  \bibinfo {author} {\bibfnamefont {M.}~\bibnamefont {Blei}}, \bibinfo {author}
  {\bibfnamefont {T.}~\bibnamefont {Taniguchi}}, \bibinfo {author}
  {\bibfnamefont {K.}~\bibnamefont {Watanabe}}, \bibinfo {author}
  {\bibfnamefont {S.}~\bibnamefont {Tongay}}, \bibinfo {author} {\bibfnamefont
  {A.}~\bibnamefont {Zettl}}, \bibinfo {author} {\bibfnamefont {M.~F.}\
  \bibnamefont {Crommie}},\ and\ \bibinfo {author} {\bibfnamefont
  {F.}~\bibnamefont {Wang}},\ }\bibfield  {title} {\bibinfo {title} {Imaging
  two-dimensional generalized {{Wigner}} crystals},\ }\href
  {https://doi.org/10.1038/s41586-021-03874-9} {\bibfield  {journal} {\bibinfo
  {journal} {Nature}\ }\textbf {\bibinfo {volume} {597}},\ \bibinfo {pages}
  {650} (\bibinfo {year} {2021}{\natexlab{a}})}\BibitemShut {NoStop}%
\bibitem [{\citenamefont {Li}\ \emph {et~al.}(2021{\natexlab{b}})\citenamefont
  {Li}, \citenamefont {Jiang}, \citenamefont {Shen}, \citenamefont {Zhang},
  \citenamefont {Li}, \citenamefont {Tao}, \citenamefont {Devakul},
  \citenamefont {Watanabe}, \citenamefont {Taniguchi}, \citenamefont {Fu},
  \citenamefont {Shan},\ and\ \citenamefont {Mak}}]{li2021quantum}%
  \BibitemOpen
  \bibfield  {author} {\bibinfo {author} {\bibfnamefont {T.}~\bibnamefont
  {Li}}, \bibinfo {author} {\bibfnamefont {S.}~\bibnamefont {Jiang}}, \bibinfo
  {author} {\bibfnamefont {B.}~\bibnamefont {Shen}}, \bibinfo {author}
  {\bibfnamefont {Y.}~\bibnamefont {Zhang}}, \bibinfo {author} {\bibfnamefont
  {L.}~\bibnamefont {Li}}, \bibinfo {author} {\bibfnamefont {Z.}~\bibnamefont
  {Tao}}, \bibinfo {author} {\bibfnamefont {T.}~\bibnamefont {Devakul}},
  \bibinfo {author} {\bibfnamefont {K.}~\bibnamefont {Watanabe}}, \bibinfo
  {author} {\bibfnamefont {T.}~\bibnamefont {Taniguchi}}, \bibinfo {author}
  {\bibfnamefont {L.}~\bibnamefont {Fu}}, \bibinfo {author} {\bibfnamefont
  {J.}~\bibnamefont {Shan}},\ and\ \bibinfo {author} {\bibfnamefont {K.~F.}\
  \bibnamefont {Mak}},\ }\bibfield  {title} {\bibinfo {title} {Quantum
  anomalous {{Hall}} effect from intertwined moir\'e bands},\ }\href
  {https://doi.org/10.1038/s41586-021-04171-1} {\bibfield  {journal} {\bibinfo
  {journal} {Nature}\ }\textbf {\bibinfo {volume} {600}},\ \bibinfo {pages}
  {641} (\bibinfo {year} {2021}{\natexlab{b}})}\BibitemShut {NoStop}%
\bibitem [{\citenamefont {Gu}\ \emph {et~al.}(2021)\citenamefont {Gu},
  \citenamefont {Ma}, \citenamefont {Liu}, \citenamefont {Watanabe},
  \citenamefont {Taniguchi}, \citenamefont {Hone}, \citenamefont {Shan},\ and\
  \citenamefont {Mak}}]{gu2021dipolar}%
  \BibitemOpen
  \bibfield  {author} {\bibinfo {author} {\bibfnamefont {J.}~\bibnamefont
  {Gu}}, \bibinfo {author} {\bibfnamefont {L.}~\bibnamefont {Ma}}, \bibinfo
  {author} {\bibfnamefont {S.}~\bibnamefont {Liu}}, \bibinfo {author}
  {\bibfnamefont {K.}~\bibnamefont {Watanabe}}, \bibinfo {author}
  {\bibfnamefont {T.}~\bibnamefont {Taniguchi}}, \bibinfo {author}
  {\bibfnamefont {J.~C.}\ \bibnamefont {Hone}}, \bibinfo {author}
  {\bibfnamefont {J.}~\bibnamefont {Shan}},\ and\ \bibinfo {author}
  {\bibfnamefont {K.~F.}\ \bibnamefont {Mak}},\ }\bibfield  {title} {\bibinfo
  {title} {Dipolar excitonic insulator in a moire lattice},\ }\href
  {http://arxiv.org/abs/2108.06588} {\bibfield  {journal} {\bibinfo  {journal}
  {arXiv:2108.06588}\ } (\bibinfo {year} {2021})}\BibitemShut {NoStop}%
\bibitem [{\citenamefont {Zhao}\ \emph {et~al.}(2022)\citenamefont {Zhao},
  \citenamefont {Kang}, \citenamefont {Li}, \citenamefont {Tschirhart},
  \citenamefont {Redekop}, \citenamefont {Watanabe}, \citenamefont {Taniguchi},
  \citenamefont {Young}, \citenamefont {Shan},\ and\ \citenamefont
  {Mak}}]{zhao2022realization}%
  \BibitemOpen
  \bibfield  {author} {\bibinfo {author} {\bibfnamefont {W.}~\bibnamefont
  {Zhao}}, \bibinfo {author} {\bibfnamefont {K.}~\bibnamefont {Kang}}, \bibinfo
  {author} {\bibfnamefont {L.}~\bibnamefont {Li}}, \bibinfo {author}
  {\bibfnamefont {C.}~\bibnamefont {Tschirhart}}, \bibinfo {author}
  {\bibfnamefont {E.}~\bibnamefont {Redekop}}, \bibinfo {author} {\bibfnamefont
  {K.}~\bibnamefont {Watanabe}}, \bibinfo {author} {\bibfnamefont
  {T.}~\bibnamefont {Taniguchi}}, \bibinfo {author} {\bibfnamefont
  {A.}~\bibnamefont {Young}}, \bibinfo {author} {\bibfnamefont
  {J.}~\bibnamefont {Shan}},\ and\ \bibinfo {author} {\bibfnamefont {K.~F.}\
  \bibnamefont {Mak}},\ }\href {https://doi.org/10.48550/arXiv.2207.02312}
  {\bibinfo {title} {Realization of the {{Haldane Chern}} insulator in a
  moir\textbackslash 'e lattice}} (\bibinfo {year} {2022})\BibitemShut
  {NoStop}%
\bibitem [{\citenamefont {Xie}\ \emph {et~al.}(2022{\natexlab{a}})\citenamefont
  {Xie}, \citenamefont {Zhang}, \citenamefont {Hu}, \citenamefont {Mak},\ and\
  \citenamefont {Law}}]{xie2022valleypolarized}%
  \BibitemOpen
  \bibfield  {author} {\bibinfo {author} {\bibfnamefont {Y.-M.}\ \bibnamefont
  {Xie}}, \bibinfo {author} {\bibfnamefont {C.-P.}\ \bibnamefont {Zhang}},
  \bibinfo {author} {\bibfnamefont {J.-X.}\ \bibnamefont {Hu}}, \bibinfo
  {author} {\bibfnamefont {K.~F.}\ \bibnamefont {Mak}},\ and\ \bibinfo {author}
  {\bibfnamefont {K.~T.}\ \bibnamefont {Law}},\ }\bibfield  {title} {\bibinfo
  {title} {Valley-{{Polarized Quantum Anomalous Hall State}} in {{Moir\'e}}
  {${\mathrm{MoTe}}_{2}/{\mathrm{WSe}}_{2}$} {{Heterobilayers}}},\ }\href
  {https://doi.org/10.1103/PhysRevLett.128.026402} {\bibfield  {journal}
  {\bibinfo  {journal} {Phys. Rev. Lett.}\ }\textbf {\bibinfo {volume} {128}},\
  \bibinfo {pages} {026402} (\bibinfo {year} {2022}{\natexlab{a}})}\BibitemShut
  {NoStop}%
\bibitem [{\citenamefont {Xie}\ \emph {et~al.}(2022{\natexlab{b}})\citenamefont
  {Xie}, \citenamefont {Zhang},\ and\ \citenamefont
  {Law}}]{xie2022topological}%
  \BibitemOpen
  \bibfield  {author} {\bibinfo {author} {\bibfnamefont {Y.-M.}\ \bibnamefont
  {Xie}}, \bibinfo {author} {\bibfnamefont {C.-P.}\ \bibnamefont {Zhang}},\
  and\ \bibinfo {author} {\bibfnamefont {K.~T.}\ \bibnamefont {Law}},\ }\href
  {https://doi.org/10.48550/arXiv.2206.11666} {\bibinfo {title} {Topological
  {$p_x+ip_y$} inter-valley coherent state in {{Moir}}\textbackslash 'e
  {{MoTe}}{$_2$}/{{WSe}}{$_2$} heterobilayers}} (\bibinfo {year}
  {2022}{\natexlab{b}})\BibitemShut {NoStop}%
\bibitem [{\citenamefont {Tao}\ \emph {et~al.}(2022)\citenamefont {Tao},
  \citenamefont {Shen}, \citenamefont {Jiang}, \citenamefont {Li},
  \citenamefont {Li}, \citenamefont {Ma}, \citenamefont {Zhao}, \citenamefont
  {Hu}, \citenamefont {Pistunova}, \citenamefont {Watanabe}, \citenamefont
  {Taniguchi}, \citenamefont {Heinz}, \citenamefont {Mak},\ and\ \citenamefont
  {Shan}}]{tao2022valleycoherent}%
  \BibitemOpen
  \bibfield  {author} {\bibinfo {author} {\bibfnamefont {Z.}~\bibnamefont
  {Tao}}, \bibinfo {author} {\bibfnamefont {B.}~\bibnamefont {Shen}}, \bibinfo
  {author} {\bibfnamefont {S.}~\bibnamefont {Jiang}}, \bibinfo {author}
  {\bibfnamefont {T.}~\bibnamefont {Li}}, \bibinfo {author} {\bibfnamefont
  {L.}~\bibnamefont {Li}}, \bibinfo {author} {\bibfnamefont {L.}~\bibnamefont
  {Ma}}, \bibinfo {author} {\bibfnamefont {W.}~\bibnamefont {Zhao}}, \bibinfo
  {author} {\bibfnamefont {J.}~\bibnamefont {Hu}}, \bibinfo {author}
  {\bibfnamefont {K.}~\bibnamefont {Pistunova}}, \bibinfo {author}
  {\bibfnamefont {K.}~\bibnamefont {Watanabe}}, \bibinfo {author}
  {\bibfnamefont {T.}~\bibnamefont {Taniguchi}}, \bibinfo {author}
  {\bibfnamefont {T.~F.}\ \bibnamefont {Heinz}}, \bibinfo {author}
  {\bibfnamefont {K.~F.}\ \bibnamefont {Mak}},\ and\ \bibinfo {author}
  {\bibfnamefont {J.}~\bibnamefont {Shan}},\ }\href
  {https://doi.org/10.48550/arXiv.2208.07452} {\bibinfo {title}
  {Valley-coherent quantum anomalous {{Hall}} state in {{AB-stacked
  MoTe2}}/{{WSe2}} bilayers}} (\bibinfo {year} {2022})\BibitemShut {NoStop}%
\bibitem [{\citenamefont {Liu}\ \emph {et~al.}(2021)\citenamefont {Liu},
  \citenamefont {Taniguchi}, \citenamefont {Watanabe}, \citenamefont {Gabor},
  \citenamefont {Cui},\ and\ \citenamefont {Lui}}]{liu2021excitonic}%
  \BibitemOpen
  \bibfield  {author} {\bibinfo {author} {\bibfnamefont {E.}~\bibnamefont
  {Liu}}, \bibinfo {author} {\bibfnamefont {T.}~\bibnamefont {Taniguchi}},
  \bibinfo {author} {\bibfnamefont {K.}~\bibnamefont {Watanabe}}, \bibinfo
  {author} {\bibfnamefont {N.~M.}\ \bibnamefont {Gabor}}, \bibinfo {author}
  {\bibfnamefont {Y.-T.}\ \bibnamefont {Cui}},\ and\ \bibinfo {author}
  {\bibfnamefont {C.~H.}\ \bibnamefont {Lui}},\ }\bibfield  {title} {\bibinfo
  {title} {Excitonic and {{Valley-Polarization Signatures}} of {{Fractional
  Correlated Electronic Phases}} in a {${\mathrm{WSe}}_{2}/{\mathrm{WS}}_{2}$}
  {{Moir\'e Superlattice}}},\ }\href
  {https://doi.org/10.1103/PhysRevLett.127.037402} {\bibfield  {journal}
  {\bibinfo  {journal} {Phys. Rev. Lett.}\ }\textbf {\bibinfo {volume} {127}},\
  \bibinfo {pages} {037402} (\bibinfo {year} {2021})}\BibitemShut {NoStop}%
\bibitem [{\citenamefont {Foutty}\ \emph {et~al.}(2023)\citenamefont {Foutty},
  \citenamefont {Kometter}, \citenamefont {Devakul}, \citenamefont {Reddy},
  \citenamefont {Watanabe}, \citenamefont {Taniguchi}, \citenamefont {Fu},\
  and\ \citenamefont {Feldman}}]{foutty2023mapping}%
  \BibitemOpen
  \bibfield  {author} {\bibinfo {author} {\bibfnamefont {B.~A.}\ \bibnamefont
  {Foutty}}, \bibinfo {author} {\bibfnamefont {C.~R.}\ \bibnamefont
  {Kometter}}, \bibinfo {author} {\bibfnamefont {T.}~\bibnamefont {Devakul}},
  \bibinfo {author} {\bibfnamefont {A.~P.}\ \bibnamefont {Reddy}}, \bibinfo
  {author} {\bibfnamefont {K.}~\bibnamefont {Watanabe}}, \bibinfo {author}
  {\bibfnamefont {T.}~\bibnamefont {Taniguchi}}, \bibinfo {author}
  {\bibfnamefont {L.}~\bibnamefont {Fu}},\ and\ \bibinfo {author}
  {\bibfnamefont {B.~E.}\ \bibnamefont {Feldman}},\ }\href
  {https://doi.org/10.48550/arXiv.2304.09808} {\bibinfo {title} {Mapping
  twist-tuned multi-band topology in bilayer {{WSe}}{$_2$}}} (\bibinfo {year}
  {2023})\BibitemShut {NoStop}%
\bibitem [{\citenamefont {Zhang}\ \emph
  {et~al.}(2021{\natexlab{a}})\citenamefont {Zhang}, \citenamefont {Devakul},\
  and\ \citenamefont {Fu}}]{zhang2021spintextured}%
  \BibitemOpen
  \bibfield  {author} {\bibinfo {author} {\bibfnamefont {Y.}~\bibnamefont
  {Zhang}}, \bibinfo {author} {\bibfnamefont {T.}~\bibnamefont {Devakul}},\
  and\ \bibinfo {author} {\bibfnamefont {L.}~\bibnamefont {Fu}},\ }\bibfield
  {title} {\bibinfo {title} {Spin-textured {{Chern}} bands in {{AB-stacked}}
  transition metal dichalcogenide bilayers},\ }\bibfield  {journal} {\bibinfo
  {journal} {PNAS}\ }\textbf {\bibinfo {volume} {118}},\ \href
  {https://doi.org/10.1073/pnas.2112673118} {10.1073/pnas.2112673118} (\bibinfo
  {year} {2021}{\natexlab{a}})\BibitemShut {NoStop}%
\bibitem [{\citenamefont {Wu}\ \emph {et~al.}(2019)\citenamefont {Wu},
  \citenamefont {Lovorn}, \citenamefont {Tutuc}, \citenamefont {Martin},\ and\
  \citenamefont {MacDonald}}]{wu2019topological}%
  \BibitemOpen
  \bibfield  {author} {\bibinfo {author} {\bibfnamefont {F.}~\bibnamefont
  {Wu}}, \bibinfo {author} {\bibfnamefont {T.}~\bibnamefont {Lovorn}}, \bibinfo
  {author} {\bibfnamefont {E.}~\bibnamefont {Tutuc}}, \bibinfo {author}
  {\bibfnamefont {I.}~\bibnamefont {Martin}},\ and\ \bibinfo {author}
  {\bibfnamefont {A.~H.}\ \bibnamefont {MacDonald}},\ }\bibfield  {title}
  {\bibinfo {title} {Topological insulators in twisted transition metal
  dichalcogenide homobilayers},\ }\href
  {https://doi.org/10.1103/PhysRevLett.122.086402} {\bibfield  {journal}
  {\bibinfo  {journal} {Phys. Rev. Lett.}\ }\textbf {\bibinfo {volume} {122}},\
  \bibinfo {pages} {086402} (\bibinfo {year} {2019})}\BibitemShut {NoStop}%
\bibitem [{\citenamefont {Angeli}\ and\ \citenamefont
  {MacDonald}(2021)}]{angeli2021valley}%
  \BibitemOpen
  \bibfield  {author} {\bibinfo {author} {\bibfnamefont {M.}~\bibnamefont
  {Angeli}}\ and\ \bibinfo {author} {\bibfnamefont {A.~H.}\ \bibnamefont
  {MacDonald}},\ }\bibfield  {title} {\bibinfo {title} {{$\Gamma$} valley
  transition metal dichalcogenide moir\'e bands},\ }\href
  {https://doi.org/10.1073/pnas.2021826118} {\bibfield  {journal} {\bibinfo
  {journal} {Proceedings of the National Academy of Sciences}\ }\textbf
  {\bibinfo {volume} {118}},\ \bibinfo {pages} {e2021826118} (\bibinfo {year}
  {2021})}\BibitemShut {NoStop}%
\bibitem [{\citenamefont {Zhang}\ \emph
  {et~al.}(2021{\natexlab{b}})\citenamefont {Zhang}, \citenamefont {Liu},\ and\
  \citenamefont {Fu}}]{zhang2021electronic}%
  \BibitemOpen
  \bibfield  {author} {\bibinfo {author} {\bibfnamefont {Y.}~\bibnamefont
  {Zhang}}, \bibinfo {author} {\bibfnamefont {T.}~\bibnamefont {Liu}},\ and\
  \bibinfo {author} {\bibfnamefont {L.}~\bibnamefont {Fu}},\ }\bibfield
  {title} {\bibinfo {title} {Electronic structures, charge transfer, and charge
  order in twisted transition metal dichalcogenide bilayers},\ }\href
  {https://doi.org/10.1103/PhysRevB.103.155142} {\bibfield  {journal} {\bibinfo
   {journal} {Phys. Rev. B}\ }\textbf {\bibinfo {volume} {103}},\ \bibinfo
  {pages} {155142} (\bibinfo {year} {2021}{\natexlab{b}})}\BibitemShut
  {NoStop}%
\bibitem [{\citenamefont {Xian}\ \emph {et~al.}(2021)\citenamefont {Xian},
  \citenamefont {Claassen}, \citenamefont {Kiese}, \citenamefont {Scherer},
  \citenamefont {Trebst}, \citenamefont {Kennes},\ and\ \citenamefont
  {Rubio}}]{xian2021realization}%
  \BibitemOpen
  \bibfield  {author} {\bibinfo {author} {\bibfnamefont {L.}~\bibnamefont
  {Xian}}, \bibinfo {author} {\bibfnamefont {M.}~\bibnamefont {Claassen}},
  \bibinfo {author} {\bibfnamefont {D.}~\bibnamefont {Kiese}}, \bibinfo
  {author} {\bibfnamefont {M.~M.}\ \bibnamefont {Scherer}}, \bibinfo {author}
  {\bibfnamefont {S.}~\bibnamefont {Trebst}}, \bibinfo {author} {\bibfnamefont
  {D.~M.}\ \bibnamefont {Kennes}},\ and\ \bibinfo {author} {\bibfnamefont
  {A.}~\bibnamefont {Rubio}},\ }\bibfield  {title} {\bibinfo {title}
  {Realization of nearly dispersionless bands with strong orbital anisotropy
  from destructive interference in twisted bilayer {{MoS2}}},\ }\href
  {https://doi.org/10.1038/s41467-021-25922-8} {\bibfield  {journal} {\bibinfo
  {journal} {Nat Commun}\ }\textbf {\bibinfo {volume} {12}},\ \bibinfo {pages}
  {5644} (\bibinfo {year} {2021})}\BibitemShut {NoStop}%
\bibitem [{\citenamefont {Movva}\ \emph {et~al.}(2018)\citenamefont {Movva},
  \citenamefont {Lovorn}, \citenamefont {Fallahazad}, \citenamefont {Larentis},
  \citenamefont {Kim}, \citenamefont {Taniguchi}, \citenamefont {Watanabe},
  \citenamefont {Banerjee}, \citenamefont {MacDonald},\ and\ \citenamefont
  {Tutuc}}]{movva2018tunable}%
  \BibitemOpen
  \bibfield  {author} {\bibinfo {author} {\bibfnamefont {H.~C.~P.}\
  \bibnamefont {Movva}}, \bibinfo {author} {\bibfnamefont {T.}~\bibnamefont
  {Lovorn}}, \bibinfo {author} {\bibfnamefont {B.}~\bibnamefont {Fallahazad}},
  \bibinfo {author} {\bibfnamefont {S.}~\bibnamefont {Larentis}}, \bibinfo
  {author} {\bibfnamefont {K.}~\bibnamefont {Kim}}, \bibinfo {author}
  {\bibfnamefont {T.}~\bibnamefont {Taniguchi}}, \bibinfo {author}
  {\bibfnamefont {K.}~\bibnamefont {Watanabe}}, \bibinfo {author}
  {\bibfnamefont {S.~K.}\ \bibnamefont {Banerjee}}, \bibinfo {author}
  {\bibfnamefont {A.~H.}\ \bibnamefont {MacDonald}},\ and\ \bibinfo {author}
  {\bibfnamefont {E.}~\bibnamefont {Tutuc}},\ }\bibfield  {title} {\bibinfo
  {title} {Tunable {$\mathrm{\ensuremath{\Gamma}}\ensuremath{-}K$} {{Valley
  Populations}} in {{Hole-Doped Trilayer}} {${\mathrm{WSe}}_{2}$}},\ }\href
  {https://doi.org/10.1103/PhysRevLett.120.107703} {\bibfield  {journal}
  {\bibinfo  {journal} {Phys. Rev. Lett.}\ }\textbf {\bibinfo {volume} {120}},\
  \bibinfo {pages} {107703} (\bibinfo {year} {2018})}\BibitemShut {NoStop}%
\bibitem [{\citenamefont {Pan}\ \emph {et~al.}(2020)\citenamefont {Pan},
  \citenamefont {Wu},\ and\ \citenamefont {Das~Sarma}}]{pan2020band}%
  \BibitemOpen
  \bibfield  {author} {\bibinfo {author} {\bibfnamefont {H.}~\bibnamefont
  {Pan}}, \bibinfo {author} {\bibfnamefont {F.}~\bibnamefont {Wu}},\ and\
  \bibinfo {author} {\bibfnamefont {S.}~\bibnamefont {Das~Sarma}},\ }\bibfield
  {title} {\bibinfo {title} {Band topology, {{Hubbard}} model, {{Heisenberg}}
  model, and {{Dzyaloshinskii-Moriya}} interaction in twisted bilayer
  {${\mathrm{WSe}}_{2}$}},\ }\href
  {https://doi.org/10.1103/PhysRevResearch.2.033087} {\bibfield  {journal}
  {\bibinfo  {journal} {Phys. Rev. Research}\ }\textbf {\bibinfo {volume}
  {2}},\ \bibinfo {pages} {033087} (\bibinfo {year} {2020})}\BibitemShut
  {NoStop}%
\bibitem [{\citenamefont {Gatti}\ \emph {et~al.}(2022)\citenamefont {Gatti},
  \citenamefont {Issing}, \citenamefont {Rademaker}, \citenamefont {Margot},
  \citenamefont {{de Jong}}, \citenamefont {{van der Molen}}, \citenamefont
  {Teyssier}, \citenamefont {Kim}, \citenamefont {Watson}, \citenamefont
  {Cacho}, \citenamefont {Dudin}, \citenamefont {Avila}, \citenamefont
  {Edwards}, \citenamefont {Paruch}, \citenamefont {Ubrig}, \citenamefont
  {{Guti{\'e}rrez-Lezama}}, \citenamefont {Morpurgo}, \citenamefont {Tamai},\
  and\ \citenamefont {Baumberger}}]{gatti2022observationa}%
  \BibitemOpen
  \bibfield  {author} {\bibinfo {author} {\bibfnamefont {G.}~\bibnamefont
  {Gatti}}, \bibinfo {author} {\bibfnamefont {J.}~\bibnamefont {Issing}},
  \bibinfo {author} {\bibfnamefont {L.}~\bibnamefont {Rademaker}}, \bibinfo
  {author} {\bibfnamefont {F.}~\bibnamefont {Margot}}, \bibinfo {author}
  {\bibfnamefont {T.~A.}\ \bibnamefont {{de Jong}}}, \bibinfo {author}
  {\bibfnamefont {S.~J.}\ \bibnamefont {{van der Molen}}}, \bibinfo {author}
  {\bibfnamefont {J.}~\bibnamefont {Teyssier}}, \bibinfo {author}
  {\bibfnamefont {T.~K.}\ \bibnamefont {Kim}}, \bibinfo {author} {\bibfnamefont
  {M.~D.}\ \bibnamefont {Watson}}, \bibinfo {author} {\bibfnamefont
  {C.}~\bibnamefont {Cacho}}, \bibinfo {author} {\bibfnamefont
  {P.}~\bibnamefont {Dudin}}, \bibinfo {author} {\bibfnamefont
  {J.}~\bibnamefont {Avila}}, \bibinfo {author} {\bibfnamefont {K.~C.}\
  \bibnamefont {Edwards}}, \bibinfo {author} {\bibfnamefont {P.}~\bibnamefont
  {Paruch}}, \bibinfo {author} {\bibfnamefont {N.}~\bibnamefont {Ubrig}},
  \bibinfo {author} {\bibfnamefont {I.}~\bibnamefont {{Guti{\'e}rrez-Lezama}}},
  \bibinfo {author} {\bibfnamefont {A.}~\bibnamefont {Morpurgo}}, \bibinfo
  {author} {\bibfnamefont {A.}~\bibnamefont {Tamai}},\ and\ \bibinfo {author}
  {\bibfnamefont {F.}~\bibnamefont {Baumberger}},\ }\href
  {http://arxiv.org/abs/2211.01192} {\bibinfo {title} {Observation of flat
  {$\Gamma$} moir\textbackslash 'e bands in twisted bilayer {{WSe}}{$_2$}}}
  (\bibinfo {year} {2022})\BibitemShut {NoStop}%
\bibitem [{\citenamefont {Pei}\ \emph {et~al.}(2022)\citenamefont {Pei},
  \citenamefont {Wang}, \citenamefont {Zhou}, \citenamefont {He}, \citenamefont
  {An}, \citenamefont {He}, \citenamefont {Chen}, \citenamefont {Li},
  \citenamefont {Wei}, \citenamefont {Liang}, \citenamefont {Avila},
  \citenamefont {Dudin}, \citenamefont {Kandyba}, \citenamefont {Giampietri},
  \citenamefont {Cattelan}, \citenamefont {Barinov}, \citenamefont {Liu},
  \citenamefont {Liu}, \citenamefont {Weng}, \citenamefont {Wang},
  \citenamefont {Xue},\ and\ \citenamefont {Chen}}]{pei2022observation}%
  \BibitemOpen
  \bibfield  {author} {\bibinfo {author} {\bibfnamefont {D.}~\bibnamefont
  {Pei}}, \bibinfo {author} {\bibfnamefont {B.}~\bibnamefont {Wang}}, \bibinfo
  {author} {\bibfnamefont {Z.}~\bibnamefont {Zhou}}, \bibinfo {author}
  {\bibfnamefont {Z.}~\bibnamefont {He}}, \bibinfo {author} {\bibfnamefont
  {L.}~\bibnamefont {An}}, \bibinfo {author} {\bibfnamefont {S.}~\bibnamefont
  {He}}, \bibinfo {author} {\bibfnamefont {C.}~\bibnamefont {Chen}}, \bibinfo
  {author} {\bibfnamefont {Y.}~\bibnamefont {Li}}, \bibinfo {author}
  {\bibfnamefont {L.}~\bibnamefont {Wei}}, \bibinfo {author} {\bibfnamefont
  {A.}~\bibnamefont {Liang}}, \bibinfo {author} {\bibfnamefont
  {J.}~\bibnamefont {Avila}}, \bibinfo {author} {\bibfnamefont
  {P.}~\bibnamefont {Dudin}}, \bibinfo {author} {\bibfnamefont
  {V.}~\bibnamefont {Kandyba}}, \bibinfo {author} {\bibfnamefont
  {A.}~\bibnamefont {Giampietri}}, \bibinfo {author} {\bibfnamefont
  {M.}~\bibnamefont {Cattelan}}, \bibinfo {author} {\bibfnamefont
  {A.}~\bibnamefont {Barinov}}, \bibinfo {author} {\bibfnamefont
  {Z.}~\bibnamefont {Liu}}, \bibinfo {author} {\bibfnamefont {J.}~\bibnamefont
  {Liu}}, \bibinfo {author} {\bibfnamefont {H.}~\bibnamefont {Weng}}, \bibinfo
  {author} {\bibfnamefont {N.}~\bibnamefont {Wang}}, \bibinfo {author}
  {\bibfnamefont {J.}~\bibnamefont {Xue}},\ and\ \bibinfo {author}
  {\bibfnamefont {Y.}~\bibnamefont {Chen}},\ }\bibfield  {title} {\bibinfo
  {title} {Observation of {{$\Gamma$}} -{{Valley Moir\'e Bands}} and {{Emergent
  Hexagonal Lattice}} in {{Twisted Transition Metal Dichalcogenides}}},\ }\href
  {https://doi.org/10.1103/PhysRevX.12.021065} {\bibfield  {journal} {\bibinfo
  {journal} {Phys. Rev. X}\ }\textbf {\bibinfo {volume} {12}},\ \bibinfo
  {pages} {021065} (\bibinfo {year} {2022})}\BibitemShut {NoStop}%
\bibitem [{\citenamefont {Foutty}\ \emph {et~al.}(2022)\citenamefont {Foutty},
  \citenamefont {Yu}, \citenamefont {Devakul}, \citenamefont {Kometter},
  \citenamefont {Zhang}, \citenamefont {Watanabe}, \citenamefont {Taniguchi},
  \citenamefont {Fu},\ and\ \citenamefont {Feldman}}]{foutty2022tunable}%
  \BibitemOpen
  \bibfield  {author} {\bibinfo {author} {\bibfnamefont {B.~A.}\ \bibnamefont
  {Foutty}}, \bibinfo {author} {\bibfnamefont {J.}~\bibnamefont {Yu}}, \bibinfo
  {author} {\bibfnamefont {T.}~\bibnamefont {Devakul}}, \bibinfo {author}
  {\bibfnamefont {C.~R.}\ \bibnamefont {Kometter}}, \bibinfo {author}
  {\bibfnamefont {Y.}~\bibnamefont {Zhang}}, \bibinfo {author} {\bibfnamefont
  {K.}~\bibnamefont {Watanabe}}, \bibinfo {author} {\bibfnamefont
  {T.}~\bibnamefont {Taniguchi}}, \bibinfo {author} {\bibfnamefont
  {L.}~\bibnamefont {Fu}},\ and\ \bibinfo {author} {\bibfnamefont {B.~E.}\
  \bibnamefont {Feldman}},\ }\bibfield  {title} {\bibinfo {title} {Tunable spin
  and valley excitations of correlated insulators in {$\Gamma$}-valley
  moir\textbackslash 'e bands},\ }\bibfield  {journal} {\bibinfo  {journal}
  {arXiv.2206.10631}\ }\href {https://doi.org/10.48550/arXiv.2206.10631}
  {10.48550/arXiv.2206.10631} (\bibinfo {year} {2022})\BibitemShut {NoStop}%
\bibitem [{\citenamefont {Mak}\ \emph {et~al.}(2010)\citenamefont {Mak},
  \citenamefont {Lee}, \citenamefont {Hone}, \citenamefont {Shan},\ and\
  \citenamefont {Heinz}}]{mak2010atomically}%
  \BibitemOpen
  \bibfield  {author} {\bibinfo {author} {\bibfnamefont {K.~F.}\ \bibnamefont
  {Mak}}, \bibinfo {author} {\bibfnamefont {C.}~\bibnamefont {Lee}}, \bibinfo
  {author} {\bibfnamefont {J.}~\bibnamefont {Hone}}, \bibinfo {author}
  {\bibfnamefont {J.}~\bibnamefont {Shan}},\ and\ \bibinfo {author}
  {\bibfnamefont {T.~F.}\ \bibnamefont {Heinz}},\ }\bibfield  {title} {\bibinfo
  {title} {Atomically {{Thin}} {${\mathrm{MoS}}_{2}$}: {{A New Direct-Gap
  Semiconductor}}},\ }\href {https://doi.org/10.1103/PhysRevLett.105.136805}
  {\bibfield  {journal} {\bibinfo  {journal} {Phys. Rev. Lett.}\ }\textbf
  {\bibinfo {volume} {105}},\ \bibinfo {pages} {136805} (\bibinfo {year}
  {2010})}\BibitemShut {NoStop}%
\bibitem [{\citenamefont {Xu}\ \emph {et~al.}(2014)\citenamefont {Xu},
  \citenamefont {Yao}, \citenamefont {Xiao},\ and\ \citenamefont
  {Heinz}}]{xu2014spin}%
  \BibitemOpen
  \bibfield  {author} {\bibinfo {author} {\bibfnamefont {X.}~\bibnamefont
  {Xu}}, \bibinfo {author} {\bibfnamefont {W.}~\bibnamefont {Yao}}, \bibinfo
  {author} {\bibfnamefont {D.}~\bibnamefont {Xiao}},\ and\ \bibinfo {author}
  {\bibfnamefont {T.~F.}\ \bibnamefont {Heinz}},\ }\bibfield  {title} {\bibinfo
  {title} {Spin and pseudospins in layered transition metal dichalcogenides},\
  }\href {https://doi.org/10.1038/nphys2942} {\bibfield  {journal} {\bibinfo
  {journal} {Nature Phys}\ }\textbf {\bibinfo {volume} {10}},\ \bibinfo {pages}
  {343} (\bibinfo {year} {2014})}\BibitemShut {NoStop}%
\bibitem [{\citenamefont {Fang}\ \emph {et~al.}(2015)\citenamefont {Fang},
  \citenamefont {Kuate~Defo}, \citenamefont {Shirodkar}, \citenamefont {Lieu},
  \citenamefont {Tritsaris},\ and\ \citenamefont {Kaxiras}}]{fang2015initio}%
  \BibitemOpen
  \bibfield  {author} {\bibinfo {author} {\bibfnamefont {S.}~\bibnamefont
  {Fang}}, \bibinfo {author} {\bibfnamefont {R.}~\bibnamefont {Kuate~Defo}},
  \bibinfo {author} {\bibfnamefont {S.~N.}\ \bibnamefont {Shirodkar}}, \bibinfo
  {author} {\bibfnamefont {S.}~\bibnamefont {Lieu}}, \bibinfo {author}
  {\bibfnamefont {G.~A.}\ \bibnamefont {Tritsaris}},\ and\ \bibinfo {author}
  {\bibfnamefont {E.}~\bibnamefont {Kaxiras}},\ }\bibfield  {title} {\bibinfo
  {title} {Ab initio tight-binding {{Hamiltonian}} for transition metal
  dichalcogenides},\ }\href {https://doi.org/10.1103/PhysRevB.92.205108}
  {\bibfield  {journal} {\bibinfo  {journal} {Phys. Rev. B}\ }\textbf {\bibinfo
  {volume} {92}},\ \bibinfo {pages} {205108} (\bibinfo {year}
  {2015})}\BibitemShut {NoStop}%
\bibitem [{\citenamefont {He}\ \emph {et~al.}(2014)\citenamefont {He},
  \citenamefont {Hummer},\ and\ \citenamefont {Franchini}}]{he2014stacking}%
  \BibitemOpen
  \bibfield  {author} {\bibinfo {author} {\bibfnamefont {J.}~\bibnamefont
  {He}}, \bibinfo {author} {\bibfnamefont {K.}~\bibnamefont {Hummer}},\ and\
  \bibinfo {author} {\bibfnamefont {C.}~\bibnamefont {Franchini}},\ }\bibfield
  {title} {\bibinfo {title} {Stacking effects on the electronic and optical
  properties of bilayer transition metal dichalcogenides
  {${\mathrm{MoS}}_{2}$}, {${\mathrm{MoSe}}_{2}$}, {${\mathrm{WS}}_{2}$}, and
  {${\mathrm{WSe}}_{2}$}},\ }\href {https://doi.org/10.1103/PhysRevB.89.075409}
  {\bibfield  {journal} {\bibinfo  {journal} {Phys. Rev. B}\ }\textbf {\bibinfo
  {volume} {89}},\ \bibinfo {pages} {075409} (\bibinfo {year}
  {2014})}\BibitemShut {NoStop}%
\bibitem [{\citenamefont {Jung}\ \emph {et~al.}(2014)\citenamefont {Jung},
  \citenamefont {Raoux}, \citenamefont {Qiao},\ and\ \citenamefont
  {MacDonald}}]{jung2014initio}%
  \BibitemOpen
  \bibfield  {author} {\bibinfo {author} {\bibfnamefont {J.}~\bibnamefont
  {Jung}}, \bibinfo {author} {\bibfnamefont {A.}~\bibnamefont {Raoux}},
  \bibinfo {author} {\bibfnamefont {Z.}~\bibnamefont {Qiao}},\ and\ \bibinfo
  {author} {\bibfnamefont {A.~H.}\ \bibnamefont {MacDonald}},\ }\bibfield
  {title} {\bibinfo {title} {Ab initio theory of moir\textbackslash 'e
  superlattice bands in layered two-dimensional materials},\ }\href
  {https://doi.org/10.1103/PhysRevB.89.205414} {\bibfield  {journal} {\bibinfo
  {journal} {Phys. Rev. B}\ }\textbf {\bibinfo {volume} {89}},\ \bibinfo
  {pages} {205414} (\bibinfo {year} {2014})}\BibitemShut {NoStop}%
\bibitem [{\citenamefont {Jung}\ and\ \citenamefont
  {MacDonald}(2013)}]{jung2013tightbinding}%
  \BibitemOpen
  \bibfield  {author} {\bibinfo {author} {\bibfnamefont {J.}~\bibnamefont
  {Jung}}\ and\ \bibinfo {author} {\bibfnamefont {A.~H.}\ \bibnamefont
  {MacDonald}},\ }\bibfield  {title} {\bibinfo {title} {Tight-binding model for
  graphene {$\ensuremath{\pi}$}-bands from maximally localized {{Wannier}}
  functions},\ }\href {https://doi.org/10.1103/PhysRevB.87.195450} {\bibfield
  {journal} {\bibinfo  {journal} {Phys. Rev. B}\ }\textbf {\bibinfo {volume}
  {87}},\ \bibinfo {pages} {195450} (\bibinfo {year} {2013})}\BibitemShut
  {NoStop}%
\bibitem [{\citenamefont {Pietzsch}\ \emph {et~al.}(2001)\citenamefont
  {Pietzsch}, \citenamefont {Kubetzka}, \citenamefont {Bode},\ and\
  \citenamefont {Wiesendanger}}]{pietzsch2001observation}%
  \BibitemOpen
  \bibfield  {author} {\bibinfo {author} {\bibfnamefont {O.}~\bibnamefont
  {Pietzsch}}, \bibinfo {author} {\bibfnamefont {A.}~\bibnamefont {Kubetzka}},
  \bibinfo {author} {\bibfnamefont {M.}~\bibnamefont {Bode}},\ and\ \bibinfo
  {author} {\bibfnamefont {R.}~\bibnamefont {Wiesendanger}},\ }\bibfield
  {title} {\bibinfo {title} {Observation of {{Magnetic Hysteresis}} at the
  {{Nanometer Scale}} by {{Spin-Polarized Scanning Tunneling Spectroscopy}}},\
  }\href {https://doi.org/10.1126/science.1060513} {\bibfield  {journal}
  {\bibinfo  {journal} {Science}\ }\textbf {\bibinfo {volume} {292}},\ \bibinfo
  {pages} {2053} (\bibinfo {year} {2001})}\BibitemShut {NoStop}%
\end{thebibliography}%

\appendix

\section{Band structure calculation and parameter estimate}\label{app:A}
\para To show that criteria (1) and (3) introduced in the main text can be satisfied by the tdbWSe$_2$, the first task is to obtain its band structure.
However, it is not easy to derive the band structure of tdbWSe$_2$ directly at the atomic level because the emergent moir\'e superlattice contains up to ten thousand atoms even within just one moir\'e unit cell. One way to approximate the band structure of the twisted system is to use a continuum model with parameters fitted according to the untwisted system in the high-symmetry stacking configurations (MM, MX, and XM) that appear in the moir\'e unit cell of the twisted system~\cite{jung2014initio}. We will provide the details later in this section.
Therefore, in this section, we will first discuss the method to obtain the band structure of untwisted systems.

We calculate the band structure of the untwisted system using the \textit{ab initio} tight-binding Hamiltonian~\cite{fang2015initio}. The logic is to first construct a tight-binding model for the monolayer by considering five $d$ orbitals of one tungsten atom and six $p$ orbitals of the two selenium atoms in the monolayer unit cell, and including spin-orbit coupling over the entire monolayer Brillouin zone (BZ) (i.e., not just the small mBZ).  It is called \textit{ab initio} because hopping parameters are evaluated from the density functional theory.
After constructing the tight-binding model for the monolayer, we can model the interlayer tunneling between the adjacent layers by considering the hopping between the two selenium atoms on the adjacent layers. Combining these considerations, we can construct the Hamiltonian for the (untwisted) double bilayer WSe$_2$.
Finally, we diagonalize the matrix of size 88 by 88 ($11$ orbitals $\times$ 2 spins $\times$ 4 layers) at each momentum to obtain the band structure $E_r(k)$, where $r=\{\text{MM}, \text{XM}, \text{MX}\}$ represents the high-symmetry stacking configuration, and $k$ is the momentum in the entire monolayer BZ.
For the full recipe and the numerical details, we refer to Ref.~\onlinecite{fang2015initio}.

Our goal is to show that tdbWSe$_2$ can satisfy criteria (1) and (3) so that a honeycomb lattice with both sublattice and SU(2) spin rotation symmetries emerges at the moir\'e scale. 
That is to say, we require $E_{r}(\Gamma)>E_{r}(K)$ for $r=$MM and MX/XM, and $E_{\text{MX}}(\Gamma)=E_{\text{XM}}(\Gamma)>E_{\text{MM}}(\Gamma)$. Obtaining these energies from the model above requires us to make a choice for the interlayer distances in the tdbWSe$_2$ structure, which are unknown parameters. Our main goal is to demonstrate the possibility and the tunability of the emergent honeycomb lattice from tdbWSe$_2$ and estimate the relevant energy scale. We will not attempt to calculate the relaxed interlayer distances. Instead, we will make assumptions on them based on the following reasoning.
In light of the $C_{2y}$ symmetry which flips the layers, the distance $d_{12}$ between the first and second layer should be the same as the distance $d_{34}$ between the third and fourth layer. 
For the distance between the central two layers $d_{23}$, because the MM stacking corresponds to the case where all tungsten atoms (also for selenium atoms) on both layers are stacked with perfect alignment, respectively, the repulsion from the interlayer interaction should be the strongest. Therefore, the value of $d_{23}$ in the MM stacking configuration should be larger than $d_{12}$ and $d_{23}$ in the MX/XM stacking configuration.
With these considerations, we perturb the interlayer distances $d_{12}$, $d_{23}$ for the MM stacking, and $d_{23}$ for the MX/XM stacking, around the interlayer distance from the bulk WSe$_2$ data, $\sim$ 6.5 \AA{}~\cite{he2014stacking}, and find that the system can satisfy the energetics hierarchy in criteria (1) and (3) in a large parameter space around $d_{12}$=6 \AA{}. 
As an example that passes criteria (1) and (3), we choose the following interlayer distances $(d_{12},d_{23},d_{34})$=(6,6.25,6) \AA{} for MM stacking and  $(d_{12},d_{23},d_{34})$=(6,5.8,6)\AA{} for MX and XM stacking for the estimation of the parameters in the continuum model. 
Note that this choice has smaller layer distances than the bulk data. In principle, one can consider adding pressure to reduce the interlayer distance if needed.
{For example, a range of $d_{23}$ from 6 to 6.5 \AA{} (from 5.7 to 6.3 \AA{}) for MM (MX and XM) stacking with $d_{23}=5.8$ \AA{} ($d_{23}=6.25$ \AA{}) for MX and XM (MM) stacking fixed can also satisfy the both criteria.} 

\begin{figure}[ht]
    \centering
    \includegraphics[width=3.4in]{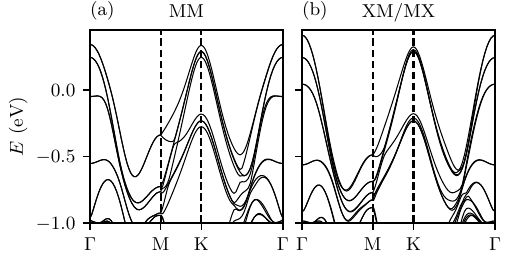}
    \caption{The valence band of tdbWSe$_2$ in the monolayer Brillouin zone for (a) MM stacking configuration and (b) MX/XM stacking configuration}
    \label{fig:Abinitio}
\end{figure}

Now, we present the valance band structure for the untwisted MM stacking in Fig.~\ref{fig:Abinitio}(a), and untwisted MX/XM stacking in Fig.~\ref{fig:Abinitio}(b). The corresponding valence band edge energies at the $\Gamma$ and $K$ valley for these two stackings are listed in Table~\ref{tab:energies}, which validate the two criteria.

To model the twisted moir\'e system from the untwisted system, 
we can represent the energy at the high-symmetry-stacking sites in one moir\'e unit cell using the energies in an untwisted system with the same type of stacking configuration. Then we interpolate them using up to the first harmonics of the moir\'e periodicity~\cite{jung2014initio}.
Therefore, by extracting the 4 topmost valence bands at the $\Gamma$-valley for the untwisted systems in the MM and MX/XM stacking configurations, we can estimate the unknown parameters in the moir\'e continuum Hamiltonian Eq.~\eqref{eq:H_4L} in the twisted system.
To do that, we first will drop the kinetic term in Eq.~\eqref{eq:H_4L}, as it is near $\Gamma$ valley, and then diagonalize the potential term with $r=$MM and $r=$MX/XM, respectively, which gives four discrete energies for each site. The fitting procedure is to adjust these four energies at $r=$MM or $r=$MX/XM to match the energies obtained from the four valence bands in the \emph{ab initio} tight-binding Hamiltonian at $\Gamma$ valley, i.e., the first two columns in Table~\ref{tab:energies}.

\begin{table}[ht]
    \begin{ruledtabular}
        \caption{First four valence band (from the top to bottom) energies (eV) at $\Gamma$ and $K$-valley for MM and MX/XM stackings }
        \begin{tabular}{lcccr}
            Band  & $E_{\text{MM}}(\Gamma)$  &  $E_{\text{MX/XM}}(\Gamma)$  & $E_{\text{MM}}(K)$  & $E_{\text{MX/XM}}(K)$ \\
            \hline
            0 & 0.341112 & 0.410080 & 0.334569 & 0.323034 \\
            1 & 0.244373 & 0.242830 & 0.286263 & 0.289225 \\
            2 & -0.050321 & 0.042969 & -0.182268 & -0.179475 \\
            3 & -0.553378 & -0.564056 & -0.232956 & -0.232795
        \end{tabular}
    \label{tab:energies}
    \end{ruledtabular}
\end{table}

\section{Phase competition in weak interaction limit}\label{app:B}
\para In this appendix, we present the analytical mean-field study of the Fermi surface instabilities in our extended Hubbard model (with the in-plane magnetic field) in the weak-coupling limit $U_{0,1}\ll t_1$ at hole filling $\nu=2$. We find that, at zero temperature, dominant instabilities include both the canted N\'eel AF order and the canted $\sqrt{3}\times \sqrt{3}$ SDW order. Their competition is controlled by the ratio $U_1/U_0$. We also discuss the relationship between our weak-coupling analysis and the Hartree-Fock phase diagrams in Fig. \ref{fig:phasediagram} obtained in the stronger coupling regime $U_0  \gtrsim t_1$. 

We start with the noninteracting part of the Hamiltonian $H^0=H_{\text{h}}+H_{\text{Z}}$ containing the hopping term and the Zeeman energy. At hole filling $\nu=2$, for small $t_2$, $h_x$ compared to $t_1$, the Fermi surfaces are centered at the $\kappa$ and $\kappa'$ points of the mBZ. We expand the dispersion around  the $\kappa$/$\kappa'$ point:
\begin{equation}
    \begin{split}        H^0_{\kappa,\sigma}&=\sum_{\bm{k}}\vec{c}_{\kappa,\sigma}^{\, \dagger}(\bm{k}) \left( h_x \sigma\Lambda^0+ v_F \vec{\Lambda} \cdot \bm{k} \right) \vec{c}_{\kappa,\sigma}(\bm{k}),\\
        H^0_{\kappa',\sigma}&=\sum_{\bm{k}}\vec{c}_{\kappa',\sigma}^{\, \dagger}(\bm{k}) \left( h_x \sigma \Lambda^0- v_F \vec{\Lambda}^* \cdot \bm{k} \right) \vec{c}_{\kappa',\sigma}(\bm{k}).
    \end{split}
    \label{eq:mDiracDisperson}
\end{equation}

Here, ${\bm k}$ is the momentum measured from the $\kappa$/$\kappa'$ point. The dispersion is expanded to the linear order of ${\bm k}$, which suffices to study the Fermi surface instabilities.  The Fermi velocity $v_F=\sqrt{3}/{2} \abs{t_1} a_M$, which is $\sim2.2\times10^4$m/s at a twist angle of $\theta = 2^\circ$. $\vec{c}_{\tau,\sigma}=\left[ c_{\tau,A,\sigma}, c_{\tau,B,\sigma}\right]^\intercal$ is the fermionic spinor in the sublattice space at valley $\tau= \kappa/\kappa'$ with spin $\sigma=\uparrow, \downarrow$ along the in-plane field direction $\hat{x}$. $\vec{\Lambda}$ comprises the Pauli-X and Y matrices acting on the sublattice space. Note that $t_2$'s contribution only starts to appear in the order-${\bm k}^2$ terms of the dispersion. This contribution is independent of valley, spin, and sublattice. Hence, a small $t_2$ is unimportant for the Fermi surface instability in the weak-coupling limit. 

At non-zero $h_x$, the Fermi surfaces of the spin-up fermions are perfectly nested with those of the spin-down fermions. Consequently, particle-hole instabilities towards magnetic ground states with spontaneous hybridization between the two spin species are expected. The magnetic orders in these ground states spontaneously break the U(1) spin rotation symmetry in the $yz$ plane (perpendicular to the in-plane field direction). In the limit $h_x \rightarrow 0$, the instabilities towards such magnetically-ordered ground states vanish along with the Fermi surfaces in the noninteracting band structure. Hence, these magnetic orders, to be discussed in detail below, are dubbed {\it in-plane-field-induced} magnetic orders. 

At hole filling $\nu=2$ (and with finite $h_x > 0$), we begin our analysis of possible instabilities by writing down the noninteracting bands crossing the Fermi level: 
\begin{equation}
    \begin{split}
        H_{\tau} = \sum_{\bm{k}} E_{+,\downarrow}(\bm{k}) c_{\tau,+,\downarrow}^\dagger(\bm{k}) c_{\tau,+,\downarrow}(\bm{k})\\
        + E_{-,\uparrow}(\bm{k}) c_{\tau,-,\downarrow}^\dagger(\bm{k}) c_{\tau,-,\downarrow}(\bm{k}),
    \end{split}
\end{equation}
for the two valleys $\tau =\kappa,\kappa'$ and the two spin species $\sigma=\uparrow,\downarrow$. The energies  $E_{\pm,\sigma}(\bm{k})$ are given by $E_{\pm,\sigma}(\bm{k})=h_x\sigma \pm v_F \abs{\bm{k}}$ with $\pm$ playing the role of a band index. This band index $\pm$ is locked with the spin index $\sigma$ for the bands that cross the Fermi level. The fermion operators $c_{\tau, \pm, \sigma}$ are defined as
\begin{equation}
    \begin{pmatrix}
        c_{\kappa,+,\sigma}^\dagger(\bm{k})\\
        c_{\kappa,-,\sigma}^\dagger(\bm{k})
    \end{pmatrix}=\frac{1}{\sqrt{2}}
    \begin{pmatrix}
        1 & e^{i\theta(\bm{k})}\\
        1 & -e^{i\theta(\bm{k})}\\
    \end{pmatrix}
    \begin{pmatrix}
        c_{\kappa,A,\sigma}^\dagger(\bm{k})\\
        c_{\kappa,B,\sigma}^\dagger(\bm{k})
    \end{pmatrix},
\end{equation}
and 
\begin{equation}
    \begin{pmatrix}
        c_{\kappa',+,\sigma}^\dagger(\bm{k})\\
        c_{\kappa',-,\sigma}^\dagger(\bm{k})
    \end{pmatrix}=\frac{1}{\sqrt{2}}
    \begin{pmatrix}
        -e^{i\theta(\bm{k})} & 1\\
        e^{i\theta(\bm{k})} & 1\\
    \end{pmatrix}
    \begin{pmatrix}
        c_{\kappa',A,\sigma}^\dagger(\bm{k})\\
        c_{\kappa',B,\sigma}^\dagger(\bm{k})
    \end{pmatrix},
\end{equation}
where $\theta(\bm{k})$ is the azimuth angle of $\bm{k}$. The spin-up (spin-down) fermions have two particle-like (hole-like) Fermi surface pockets, one around the $\kappa$ point and the other around the $\kappa'$ point. Hence, possible particle-hole instabilities (at weak couplings) can occur through either intervalley or intravalley hybridizations of the two spin species. 

First, we consider the intervalley hybridization associated with the mean-field order parameter $\expval{c_{\tau,-,\uparrow}^\dagger(\bm k) c_{\bar{\tau},+,\downarrow}(\bm k) }$. Here, $\bar{\tau}$ represents the opposite valley of $\tau$. The mean-field decomposition of the extended Hubbard interaction corresponding to the intervalley hybridization reads
\begin{widetext}
    \begin{equation}\label{eq:Hinter}
        \begin{split}
            H_{\text{int}}^{\text{inter}}&=-\sum_{\substack{\bm{k},\bm{k}',\tau,\sigma}} \left( \frac{U_0}{2N}\cos(\theta(\bm{k})-\theta(\bm{k}'))+\frac{3U_1}{2N}\right)\expval{c_{\bar{\tau},\bar{\sigma}}^\dagger(\bm{k}') c_{{\tau},{\sigma}}(\bm{k}')} c_{{\tau},{\sigma}}^\dagger(\bm{k}) c_{\bar{\tau},\bar{\sigma}}(\bm{k}) \\
            &+ \sum_{\substack{\bm{k},\bm{k}',\tau}} \left( \frac{U_0}{2N}\cos(\theta(\bm{k})-\theta(\bm{k}')) +  \frac{3U_1}{2N}\right) \expval{c_{{\tau},\uparrow}^\dagger(\bm{k}')  c_{\bar{\tau},{\downarrow}}(\bm{k}')} \expval{c_{\bar{\tau},\downarrow}^\dagger(\bm{k}) c_{{\tau},{\uparrow}}(\bm{k})} .
        \end{split}
    \end{equation}
\end{widetext}

Here, $\bar{\sigma}$ represents the opposite spin of $\sigma$. $N$ is the system size. The band index is suppressed in the shorthands $c_{{\tau},{\uparrow}}(\bm{k})=c_{{\tau},-,{\uparrow}}(\bm{k})$  and $c_{{\tau},{\downarrow}}(\bm{k})=c_{{\tau},+,{\downarrow}}(\bm{k})$ due to the locking of the spin and the band indices for the bands that cross the Fermi level. In the derivation of $H_{\text{int}}^{\text{inter}}$, the terms irrelevant to the intervalley hybridization of opposite spin species and the terms of linear or higher order in $k$ are dropped. Note that the first (second) term in the quadratic terms in the first line of Eq.~\eqref{eq:Hinter} only couples to the $p$-wave ($s$-wave) channel of the order parameter $\expval{c_{\tau,-,\uparrow}^\dagger(\bm k) c_{\bar{\tau},+,\downarrow}(\bm k) }$. We can consider the $s$-wave channel (with orbital angular momentum $\abs{l}=0$) and the $p$-wave channel  (with orbital angular momentum $\abs{l}=1$) separately. At the mean-field level, the saddle point equations in the $s$-wave channel only depend on $U_1$, while those in the $p$-wave channel only depend on $U_0$.

For the intravalley hybridization associated with the mean-field order parameter $\expval{c_{\tau,-,\uparrow}^\dagger ({\bm k}) c_{\tau,+,\downarrow} ({\bm k}) }$, the mean-field decomposition of interaction terms reads
\begin{widetext}
    \begin{equation}\label{eq:Hintra}
        \begin{split}
            H_{\text{int}}^{\text{intra}}&=-\frac{U_0}{2N}\sum_{\substack{\bm{k},\bm{k}',\tau,\sigma}}\left( \expval{c_{{\tau},\bar{\sigma}}^\dagger(\bm{k}') c_{{\tau},{\sigma}}(\bm{k}')} - \expval{c_{\bar{\tau},\bar{\sigma}}^\dagger(\bm{k}') c_{\bar{\tau},{\sigma}}(\bm{k}')} \right) c_{{\tau},{\sigma}}^\dagger(\bm{k}) c_{{\tau},\bar{\sigma}}(\bm{k}) \\
            &- \frac{3U_1}{2N} \sum_{\substack{\bm{k},\bm{k}',\tau,\sigma}} \cos(\theta(\bm{k})-\theta(\bm{k}')) \expval{c_{{\tau},\bar{\sigma}}^\dagger(\bm{k}') c_{{\tau},{\sigma}}(\bm{k}')} c_{\tau,\sigma}^\dagger(\bm{k}) c_{{\tau},\bar{\sigma}}(\bm{k}) \\
            &+ \frac{U_0}{2N}\sum_{\substack{\bm{k},\bm{k}'}} \left( \expval{c_{\kappa,\uparrow}^\dagger(\bm{k}') c_{\kappa,\downarrow}(\bm{k}')} - \expval{c_{\kappa',\uparrow}^\dagger(\bm{k}') c_{\kappa',\downarrow}(\bm{k}')} \right) \left( \expval{c_{\kappa,\downarrow}^\dagger(\bm{k}) c_{\kappa,\uparrow}(\bm{k})} - \expval{c_{\kappa',\downarrow}^\dagger(\bm{k}) c_{\kappa',\uparrow}(\bm{k})} \right) \\
            &+\frac{3U_1}{2N} \sum_{\substack{\bm{k},\bm{k}',\tau}}\cos(\theta(\bm{k})-\theta(\bm{k}')) \expval{c_{{\tau},{\uparrow}}^\dagger(\bm{k}') c_{{\tau},{\downarrow}}(\bm{k}')} \expval{c_{{\tau},{\downarrow}}^\dagger(\bm{k}) c_{{\tau},{\uparrow}}(\bm{k})}.
            \end{split}
        \end{equation}
\end{widetext}
Similar to the intervalley case, based on the form of $H_{\text{int}}^{\text{intra}}$, we only need to consider the $s$-wave and the $p$-wave channel of the order parameter $\expval{c_{\tau,-,\uparrow}^\dagger ({\bm k}) c_{\tau,+,\downarrow} ({\bm k}) }$. Also, it is natural to expect that the order parameters in the two valleys $\tau = \kappa,\kappa'$ share the same orbital angular momentum. With this expectation, we find that the saddle point equations in the $s$-wave channel only depend on $U_0$ while those in the $p$-wave channel only depend on $U_1$.

At zero temperature, we find the two most dominant mean-field solutions with the largest gaps opened around the Fermi surface. The corresponding orders are then expected to have the highest mean-field transition temperatures. One of these mean-field solutions is given by the intervalley hybridization in the $s$-wave channel with $\expval{c_{\kappa,-,\uparrow}^\dagger c_{\kappa',+,\downarrow}} = \expval{c_{\kappa',-,\uparrow}^\dagger c_{\kappa,+,\downarrow}} $, which gives rise to the canted $\sqrt{3}\times\sqrt{3}$ SDW order, as illustrated in Fig.~\ref{fig:phasediagram}(b). In this phase, the gap opened at the Fermi surface is given by $2h_x\exp(-\frac{2}{3U_1 \rho})$, where the density of states $\rho$ can be approximated to a constant $\frac{4 h_x}{3\pi t_1^2} $ within the energy interval between two Dirac points (at energies $-h_x$ and $h_x$ relative to the Fermi level). The other dominant mean-field solution is given by the intravalley hybridization in the $s$-wave channel with $\expval{c_{\kappa,-,\uparrow}^\dagger c_{\kappa,+,\downarrow}} = -\expval{c_{\kappa',-,\uparrow}^\dagger c_{\kappa',+,\downarrow}} $, which gives rise to the canted N\'eel AF order, as shown in Fig.~\ref{fig:phasediagram} (b) and (c). In this canted N\'eel AF phase, the gap size is given by $2h_x\exp(-\frac{1}{U_0 \rho})$.

By comparing the gaps, we expect the canted N\'eel AF phase to be the energetically favorable ground state for $U_1 < \frac{2}{3} U_0$ in the weak coupling limit. When $U_1$ increases beyond $\frac{2}{3} U_0$, the ground state goes through a first-order phase transition into the canted $\sqrt{3}\times \sqrt{3}$ SDW phase. We comment that it is challenging to extend our numerical Hartree-Fock calculation, which includes all the interaction terms, to the weak-coupling limit because of the exponentially small gap sizes. From the Hartree-Fock phase diagram with $U_0>t_1$ as shown in Fig. \ref{fig:phasediagram} (c), the sublattice-polarized phase is more energetically favorable than both the canted $\sqrt{3}\times \sqrt{3}$ SDW insulator and the canted N\'eel AF insulator when $U_1 \gtrsim \frac{1}{3}U_0 + \frac{2}{3}h_x$. However, there is no Fermi surface instability towards the sublattice-polarized insulator in the weak coupling limit. It is interesting to investigate the competition among the canted N\'eel AF insulator, the $\sqrt{3}\times \sqrt{3}$ SDW insulator, and the sublattice-polarized insulator in the regime with $U_0<t_1$, which we will leave for future study. 

\end{document}